\documentclass[aps,pra,reprint,amsmath,amssymb,amsfonts,twocolumn,nofootinbib,showpacs]{revtex4-1}

\usepackage{bm,graphicx,mathrsfs}
\usepackage{graphicx}
\usepackage{epsfig}
\usepackage{amsmath,bbm}
\usepackage{amsfonts,amssymb}
\usepackage{times}
\usepackage{dsfont}

\newcommand{\ket}[1]{|#1\rangle}
\newcommand{\bra}[1]{\langle#1|}

\usepackage{xpatch}
\makeatletter
\patchcmd{\@ssect@ltx}
    {\addcontentsline{toc}{#1}{\protect\numberline{}#8}}
    {}
    {}
    {}
\makeatother

\begin{document}

\title{Scalable Dissipative Preparation of Many-Body Entanglement}

\author{Florentin Reiter$^{1*}$, David Reeb$^{2}$, and Anders S. S\o{}rensen$^{1}$}
\affiliation{$^{1}$Niels Bohr Institute, University of Copenhagen, Blegdamsvej 17, DK-2100 Copenhagen, Denmark\\
$^{2}$Leibniz Universit\"{a}t Hannover, Appelstr. 2, 30167 Hannover, Germany
\\
$^*$Current address: Harvard University, Department of Physics, 17 Oxford Street, Cambridge, MA 02138, USA
}
\date{\today}

\begin{abstract}
We present a technique for the dissipative preparation of highly entangled multiparticle states of atoms coupled to common oscillator modes. By combining local spontaneous emission with coherent couplings we engineer many-body dissipation that drives the system from an arbitrary initial state into a Greenberger-Horne-Zeilinger state. We demonstrate that using our technique, highly entangled steady states can be prepared efficiently in a time that scales polynomially with the system size. Our protocol assumes generic couplings and will thus enable the dissipative production of multiparticle entanglement in a wide range of physical systems. As an example, we demonstrate the feasibility of our scheme in state-of-the-art trapped-ion systems.
\end{abstract}

\pacs{03.67.Bg, 42.50.Dv, 42.50.Lc, 03.67.Pp}

\maketitle

Multiparticle entanglement is an essential resource for quantum computation and information \cite{NielsenChuang}, e.g.\ in quantum error correction \cite{ShorErrorCorrection, Steane}, quantum memories \cite{Fleischhauer}, and entanglement-enhanced quantum measurement schemes \cite{Giovanetti, Toth}.
Among the most important states which exhibit genuine multiparticle entanglement are Greenberger-Horne-Zeilinger (GHZ) states
\begin{align}
\label{EqGHZ}
\ket{{\rm GHZ}} &= \frac{1}{\sqrt{2}}\left(\ket{000...0} + \ket{111...1}\right).
\end{align}
Deterministic preparation of such states has so far been performed using time-dependent unitary gates
\cite{Rauschenbeutel, SchmidtKaler, Sackett, DiCarloNeeley}, which have recently yielded impressive progress towards entangling larger numbers of qubits \cite{Leibfried, Haffner, Monz}, and feedback control schemes \cite{WisemanMilburn, Bergholm, Riste}.
These operations however suffer from quantum noise, causing decoherence and dissipation so that it remains difficult to prepare high-fidelity multiparticle entangled states with these methods \cite{DiVincenzo, Ladd}.
Recently, dissipative state preparation has been proposed as an alternative approach where the dissipative environment is actively engineered and used to prepare states relevant for quantum information and simulation \cite{Beige1, Kraus, Ticozzi1, Diehl, VWC}.
Numerous theoretical studies on the production of bipartite entangled states \cite{PCZ, PHBK, Beige2, WS, KRS, Gullans, Carr, Schuetz} have since been performed and the first experimental demonstrations \cite{Krauter, Barreiro, Lin, Shankar} have been realized.
More recently, also dissipative schemes for the generation of multipartite entangled states \cite{SchneiderMilburn, Vacanti, Weimer, Vollbrecht, CBK, KRW, Stevenson, FossFeig, Honing, Plasmons, Cormick, LCC, Ticozzi2, Rao, Pichler, Lee} have become available, e.g. for the preparation of states stabilized by local interactions \cite{Kraus, VWC, KRW}.
It has however remained a challenge to prepare in a scalable way states, like GHZ, that are highly entangled in the sense that they cannot be stabilized by local operators \cite{CBK, VWC, Ticozzi1}.

In this Letter, we extend the range of the dissipative approach by demonstrating a scalable technique for the dissipative preparation of highly entangled states of many particles. We show that, by using local  spontaneous emission as a generic source of dissipation, we can engineer nontrivial many-body dissipative interactions \cite{Ticozzi2} which are tailored to produce multiparticle GHZ states.
Our scheme is deterministic and operates by continuous optical driving from an arbitrary initial state towards the desired steady state using weak classical fields.
The preparation time of our protocol is found to exhibit a favorable polynomial scaling with the number of qubits.
In addition to our generic system-independent scheme, we describe an implementation in a system of trapped ions. 

\begin{figure}[t]
\centering
\includegraphics[width=\columnwidth]{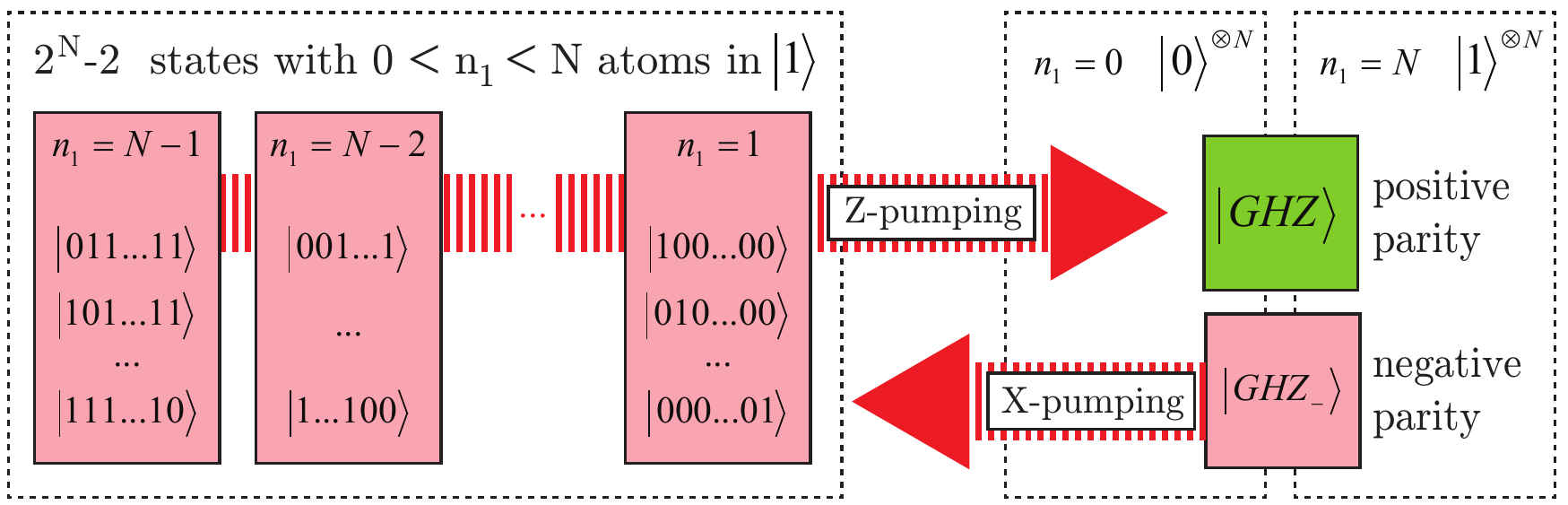}
\caption{
Protocol. Preparation of a GHZ state of $N$ qubits is realized by two operations: \textit{(i)} ``Z-Pumping'' of states with other than $0$ or $N$ atoms in state $\ket{1}$ to $\ket{0}^{\otimes N}$. \textit{(ii)} $\ket{0}^{\otimes N}$ is a superposition of $\ket{{\rm GHZ}}$ and $\ket{{\rm GHZ_-}} = (\ket{0}^{\otimes N} - \ket{1}^{\otimes N})/\sqrt{2}$ so that $\ket{\rm GHZ_-}$ needs to be removed by the parity-selective ``X-pumping''.
}
\label{FigProtocol}
\end{figure}

In our protocol, preparation of a steady GHZ state \eqref{EqGHZ} of $N$ qubits starting from an arbitrary initial state is accomplished by two simultaneous operations shown in Fig. \ref{FigProtocol}:
\textit{(i)} Pumping all states with more than zero but less than all qubits in state $\ket{1}$ ($0 < n_1 < N$) to the state $\ket{0}^{\otimes N}$, which can be written as a superposition $\ket{0}^{\otimes N} = \ket{\rm GHZ} + \ket{\rm GHZ_-}$, and \textit{(ii)} removing the GHZ state with the wrong phase, $\ket{{\rm GHZ_-}} = (\ket{0}^{\otimes N} - \ket{1}^{\otimes N})/\sqrt{2}$ from the subspace spanned by $\ket{0}^{\otimes N}$ and $\ket{1}^{\otimes N}$.
Operation \textit{(i)} is implemented such that it fulfills the main requirement of a dissipative protocol: it has to pump an exponential number of states efficiently, i.e. in polynomial time.
In principle, standard optical pumping \cite{Happer} satisfies this criterion as well, but would also erase the state $\ket{1}^{\otimes N}$, thus ruling out the possibility to prepare $\ket{\rm GHZ}$ with high fidelity. Instead, we design a new procedure which is selective in the number of atoms in $\ket{1}$. This allows us to pump only states with $0 < n_1 < N$ to states with $n_1 - 1$, and thus eventually to $\ket{0}^{\otimes N}$ ($n_1 = 0$).
We refer to this operation shown in Fig. \ref{FigProtocol} as ``Z-pumping'', since it is based on counting the number of atoms in the eigenbasis of $Z = \ket{0}\bra{0} - \ket{1} \bra{1}$.

A second operation \textit{(ii)} is required to remove the undesired $\ket{\rm GHZ_-}$ state in a continuous manner, as also illustrated in Fig. \ref{FigProtocol}. To this end, we perform a pumping process selective in the parity $\mathcal{P} = \Pi_{a=1}^N X_a$ ($X_a = \ket{1}_a\bra{0} + \ket{0}_a\bra{1}$).
Here, we apply the recipe for stabilizer pumping from Ref. \cite{CBK} to the case of the parity stabilizer, $\mathcal {P}$:
Expressed in terms of the eigenstates of $X$, $\ket{\pm} = (\ket{0} \pm \ket{1})/\sqrt{2}$, any state $\ket{\psi}$ with $\mathcal{P} \ket{\psi} = +1 \ket{\psi}$, such as $\ket{\rm GHZ}$, is a superposition of only those product states that contain an even number, $n_-$, of $\ket{-}$ qubits; e.g., for $N=3$, $\ket{\rm GHZ} = (\ket{+++}+\ket{+--}+\ket{-+-}+\ket{--+})/2$.
On the other hand, $n_-$ is odd for any state $\ket{\psi}$ with $\mathcal{P} \ket{\psi} = -1 \ket{\psi}$, such as $\ket{\rm GHZ_-}$; for $N=3$, $\ket{\rm GHZ_-} = (\ket{++-}+\ket{+-+}+\ket{-++}+\ket{---})/2$.
By pumping all states with odd $n_-$ to other states, in the following referred to as ``X-pumping'', we achieve the depumping of $\ket{\rm GHZ_-}$.

\begin{figure}[t]
\centering
\includegraphics[width=\columnwidth]{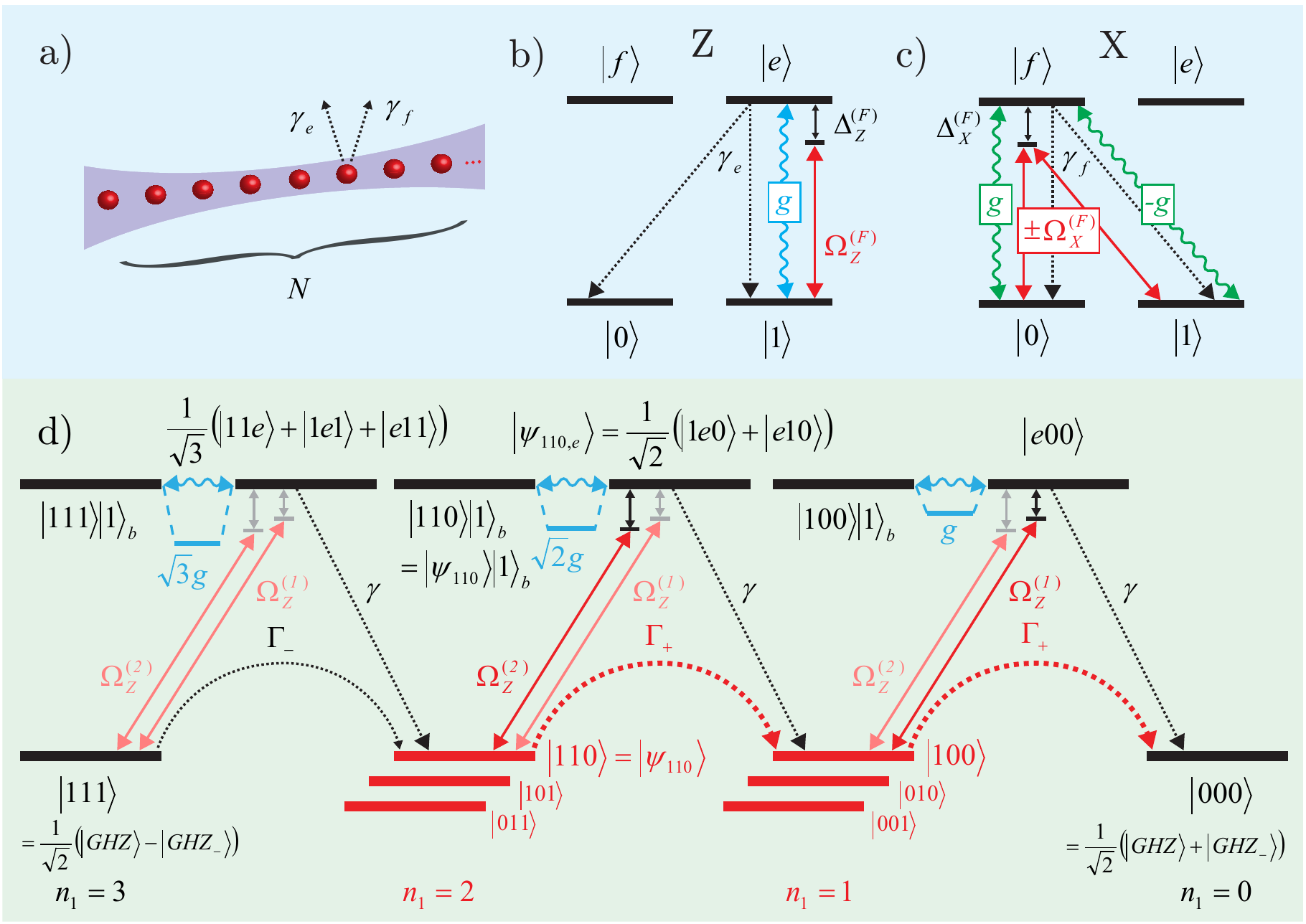}
\caption{Setup (a)-(c) and dissipative mechanism (d) for GHZ preparation, shown for $N=3$ qubits.
Setup (a). We consider a chain of $N$ sub-systems (``atoms'') with four levels, coupled to two common harmonic oscillators. As dissipative processes we assume spontaneous emission from the excited levels $\gamma_{e / f}$.
We apply coupling configurations `Z' (b), and `X' (c) consisting of $2(N-1)$ driving tones $\Omega_{Z}^{(F)}$ with detunings $\Delta_{Z}^{(F)}$ in (b), $2 \lfloor (N+1)/2 \rfloor$ tones with $\Omega_{X}^{(F)}$ and $\Delta_{X}^{(F)}$ in (c), and couplings of the atoms to the oscillator modes $g$.
(d) ``Z-pumping'' towards $\ket{0}^{\otimes N}$ using the couplings in (b). Ground states are coupled to atom-excited states by weak driving.
Depending on the number $n_1$ of atoms in $\ket{1}$, the excited states form dressed states with oscillator-excited states at energies $\pm \sqrt{n_1} g$. By applying fields with detunings $|\Delta_Z^{(F)}| = \sqrt{F} g$ for $1 \leq F \leq N-1$ (only the red-detuned fields are shown), all states except $\ket{{\rm GHZ}}$ and $\ket{{\rm GHZ_-}}$ are pumped to $\ket{0}^{\otimes N} = (\ket{{\rm GHZ}} + \ket{{\rm GHZ_-}})/\sqrt{2}$.
$\ket{\rm GHZ_-}$ is emptied by the parity-selective X-pumping as described in the text.
}
\label{FigRealization}
\end{figure}

The two operations required for our protocol can be realized using a generic setup as described next.
We assume a general system of $N$ particles (``atoms''), shown in Fig. \ref{FigRealization} (a)-(c).
Each atom supports two stable ground states $\ket{0}$ and $\ket{1}$ and two excited states $\ket{e}$ and $\ket{f}$. The atoms are driven by classical multi-tone driving fields, with identical amplitudes on all ions, as described by the Hamiltonians
\begin{align}
H_{{\rm drive}, Z}^{(F)} &= \frac{1}{2} \Omega_{\rm Z}^{(F)} e^{i \Delta_{\rm Z}^{(F)} t} \sum_{a=1}^N \ket{e}_a \bra{1} + H.c.
\label{HdriveZ}
\\
H_{{\rm drive}, X}^{(F)} &= \frac{1}{2} \Omega_{\rm X}^{(F)} e^{i \Delta_{\rm X}^{(F)} t} \sum_{a=1}^N \ket{f}_a \bra{-} + H.c.,
\label{HdriveX}
\end{align}
with strengths $\Omega_l^{(F)}$ and detunings $\Delta_l^{(F)}$, where $l=Z,X$ denotes the desired operation and $F_{\rm GHZ}$ the field tone. The transitions of the atoms are collectively coupled to two harmonic oscillator modes, $b$ and $c$, e.g., two resolved resonator modes in cavity or circuit QED \cite{Rauschenbeutel, DiCarloNeeley}, or two phononic modes in an ion trap setup \cite{SchmidtKaler, Leibfried, Haffner, Monz}, as modeled by the Hamiltonians
\begin{align}
H_{{\rm int}, Z} &= g b^\dagger \sum_{a=1}^N \ket{1}_a \bra{e} + H.c.,
\label{HintZW}
\\
H_{{\rm int}, X} &= g c^\dagger \sum_{a=1}^N \ket{-}_a \bra{f} + H.c.,
\label{HintX}
\end{align}
with $g$ being the coupling constant.
For the dissipative process we consider decay by spontaneous emission from the excited states $j$ to the ground states $i$ at a rate $\gamma_{ij}$ (with $\gamma_j = \sum_i \gamma_{ij}$), described by jump operators $L_{\gamma_{ij},a} = \sqrt{\gamma_{ij}} \ket{i}_a \bra{j}$ for $i \in \{0, 1\}$, $j \in \{e, f\}$ acting incoherently on each atom $a$.

We realize the Z- and X-pumping operations by engineering selected transitions to be driven resonantly, while suppressing others due to off-resonant driving. For Z-pumping we use the coupling configuration in Fig.\ \ref{FigRealization} (b). Here, a coupling $g$ of the oscillator $b$ to the transition $\ket{e} \leftrightarrow \ket{1}$ and a weak drive on the same transition are used to effectively ``count'' the number of atoms in state $\ket{1}$.
In Fig. \ref{FigRealization} (d), we illustrate this mechanism for $N = 3$ qubits.
The weak driving tones $\Omega_Z^{(F)}$ couple the ground states to atom-excited states. For example we consider $\ket{\psi_{110}} = \ket{110}$ which is coupled to $\ket{\psi_{110,e}} = (\ket{e10}+\ket{1e0})/\sqrt{2}$. This state is in turn coupled to an oscillator-excited state $\ket{\psi_{110}}\ket{1}_b$ by the atom-oscillator coupling. Because of constructive interference between the two terms in $\ket{\psi_{110,e}}$ this coupling has a strength of $\sqrt{2} g$.
As a consequence of the strong atom-oscillator coupling, the atom- and the oscillator-excited state form dressed states $\ket{\psi_{110,\pm}} = (\ket{\psi_{110,e}}\ket{0}_b \pm \ket{\psi_{110}}\ket{1}_b)/\sqrt{2}$ at detunings $\Delta_\pm = \pm \sqrt{2} g$, see Fig. \ref{FigRealization} (d).
Applying a weak driving field with a detuning $|\Delta_Z| = \sqrt{2} g$, one thus excites ground states with $n_1 = 2$ to excited states like $\ket{\psi_{110,e}}$. Since $\ket{e}$ is subject to spontaneous emission to $\ket{0}$ and $\ket{1}$, $\ket{\psi_{110,e}}$ decays either back to the manifold of states with $n_1 = 2$ or ``forward'' to states with $n_1 = 1$. In general, the couplings of the atom- and oscillator-excited states have a strength of $\sqrt{n_1} g$ that depends on the number $n_1$ of atoms in $\ket{1}$ so that the dressed states are shifted by $\Delta_\pm = \pm \sqrt{n_1} g$. This creates a resonance condition depending on $n_1$ of the initial state which we can use to selectively drive a manifold. Applying a drive with $|\Delta_Z| = \sqrt{1} g$, states with $n_1=1$ are thus pumped to $n_1 = 0$, and thereby to $\ket{\rm GHZ}$ and $\ket{\rm GHZ_-}$. On the other hand, with the detunings $\Delta_Z = \pm g$ and $\pm \sqrt{2}g$, the state $\ket{1}^{\otimes N}$ is excited only off-resonantly and thus decays slowly (see Fig. \ref{FigRealization} (d), so that $\ket{\rm GHZ}$ remains almost unaffected.

To realize the full Z-pumping process based on the mechanism above, we apply a weak drive consisting of $2(N-1)$ tones $\Omega_Z^{(F)}$ with detunings $\Delta_Z^{(F)} = \pm \sqrt{F} g$ ranging from $F = 1$ to $F = N-1$. This gives rise to effective decay processes \cite{SI} described by
\begin{align}
L_{\gamma_0,a,Z}^{(n_1)}
&= \sqrt{\gamma_{0,Z}^{(n_1)}} \ket{0}_a \bra{1} P_{n_1}, ~ ~ ~ (1 \leq n_1 \leq N -1).
\label{EqEngSpontEm}
\end{align}
Here, $P_{n_1}$ is the projector onto the ground states with $n_1$ atoms in state $\ket{1}$, $a$ denotes the atom subject to decay, and $\gamma_{0,Z}^{(n_1)} = \gamma_{0e}(\Omega_Z^{(n_1)}/\gamma_e)^2$ are the strongly enhanced decay rates of the states which are resonantly excited.
Each ground state with $1 \leq n_1 \leq N-1$ then decays towards one with $n_1 - 1$ by the engineered spontaneous emission in Eq. \eqref{EqEngSpontEm}.
The concatenation of these decay processes causes a continuous drift towards states with smaller $n_1$, finally ending at the state $\ket{0}^{\otimes N}$ with $n_1 = 0$. For a suitable choice of the $\Omega_Z^{(n_1)}$ \cite{SI}, the preparation time $\tau$ of $\ket{0}^{\otimes N}$ and the corresponding rate $\Gamma_+ = 1/\tau$ can then have a favorable scaling 
$\tau \propto 1/\Gamma_+ \sim \log(N)$, which is similar to optical pumping where the transition rate of each level is proportional to the number $n_1$ of its excitations.
However, as opposed to standard optical pumping, the Z-pumping is engineered such that it does not affect $\ket{1}^{\otimes N}$, and thus neither $\ket{\rm GHZ}$, since resonant excitation out of $\ket{1}^{\otimes N}$ would require the tone $F = N$. By excluding such tones from the drive the excitation out of $\ket{1}^{\otimes N}$ is off-resonant and thus much weaker.
The weak off-resonant excitation of $\ket{1}^{\otimes N}$ by the other driving tones leads only to a small leakage from $\ket{\rm GHZ}$.
Since the energy gap between the resonances and the driving tones $(\sqrt{N}-\sqrt{n_1}) g$ decreases with $N$, the leakage rate from $\ket{\rm GHZ}$ increases with the number of qubits, and can be estimated to be $\Gamma_-^Z \sim (N \gamma_e / g^2) \sum_{n_1=1}^{N-1} n_1 (\Omega_Z^{(n_1)}/(N - n_1))^2$. The resulting error can, however, be compensated by reducing the speed of the protocol by a small polynomial factor as we discuss below.

Having operation \textit{(i)} realized, we now turn to operation \textit{(ii)}, the X-pumping, which removes states with an odd number $n_-$ of atoms in $\ket{-}$.
To implement it we use a similar mechanism as was used to ``count'' $n_1$ above, except that now we need to do this in a different basis.
We achieve this operation using the coupling configuration in Fig. \ref{FigRealization} (c): Coupling the transitions from $\ket{0}$ to $\ket{f}$ and from $\ket{1}$ to $\ket{f}$ by fields of the same strength, but with the opposite phase, results in a coupling of the transition from $\ket{-}$ to $\ket{f}$.
To avoid interference with the Z-pumping, we consider a second excited state $\ket{f}$ and a second oscillator mode $c$. It is nonetheless possible to implement the described operations with a single excited level and a single oscillator mode in a stroboscopic manner, resulting in a quasi-steady state.

In the X-pumping, the transitions between the excited level $\ket{f}$ and the ground levels $\ket{0}$ and $\ket{1}$ are coherently coupled to the harmonic oscillator $c$ and excited by a multi-tone drive $\Omega_X^{(F)}$. Opposite phases on both transitions result in a coupling of the transition $\ket{f} \leftrightarrow \ket{-}$, similar to the coupling $\ket{e} \leftrightarrow \ket{1}$ in the Z configuration. In this way, we make the X-pumping selective in $n_-$ in a similar manner as the Z-pumping is selective in $n_1$:
Applying $2\lfloor (N+1)/2 \rfloor$ field tones with detunings $\Delta_X^{(F)} = \pm \sqrt{F} g$ for $F = 1,3,5, \ldots (F \leq N)$ we resonantly excite $\ket{{\rm GHZ_-}}$ to dressed states which lie at $\pm \sqrt{n_-} g$ for $n_- = 1,3,5, \ldots (n_- \leq N)$. Thereby we make $\ket{\rm GHZ_-}$ decay to random states by effective spontaneous emission with a strong rate $\Gamma_+^X \sim 2 (\Omega_X^{(F)})^2 / \gamma_f$, 
Similar to the Z-pumping, the decreasing energy gap between the dressed states gives rise to a leakage rate from GHZ, $\Gamma_-^X \sim N^2 \gamma_f \Omega_X^2 / g^2$ (using $\Omega_X^{(F)}=\Omega_X$ for odd $F$ and $\Omega_X^{(F)}=0$ for even $F$), which increases with the number of qubits $N$.

The simultaneous action of the Z- and the X-pumping prepares $\ket{\rm GHZ}$ and maintains
it as the unique steady state of the dissipative dynamics. However, since the Z-pumping is disturbed by the X-pumping, the latter has to be sufficiently weak so that the Z-pumping has a sizable probability of reaching the final state $\ket{0}^{\otimes N}$ before being subject to X-pumping; picking similar rates for the X-pumping and the total Z-pumping $\Gamma_+$, this requirement does not slow down the preparation process significantly \cite{SI}.
To find the preparation time we can consider a simple model where the rate of pumping to $\ket{0}^{\otimes N}$ is determined by $\Gamma^Z_+$, and where $\ket{\rm GHZ}$ and $\ket{\rm GHZ_-}$ are pumped out with rates $\Gamma_- = \Gamma_-^Z + \Gamma_-^X$ and $\Gamma^X_+$, respectively. Further details on the model are given in the Supplementary, but in short we find 
that the steady state fidelity $F$ is determined by the ratio of the decay out due to off-resonant excitation at the rate $\Gamma_-\sim\gamma N\Omega^2/g^2$, and the effective preparation rate $\Gamma_+ \sim \Omega^2 / \gamma$ (using $\Omega_X^{(F)} \sim \Omega/\sqrt{N \log N}$, $\Omega_Z^{(F)} \sim \Omega$ for $F\leq 2N/3$, and $\Omega_Z^{(F)} \sim \sqrt{2 (N-F)/F} \Omega$ for $F\geq 2N/3$). This gives a steady state error ${\mathcal E}=\Gamma_-/\Gamma_+\sim N\gamma^2/g^2$ which is approached exponentially in time $\sim e^{-t\Gamma_+}$.

To avoid having an increasing error with the qubit number $N$ we assume that we can control the decay rates of the excited states $\ket{e}$ and $\ket{f}$. This is for instance the case if the states are metastable states coupled to higher lying unstable states with a laser field \cite{Barreiro, Lin}.
The increase of the error with $N$ can then be compensated by having a sufficiently low decay rate $\gamma \sim g \sqrt{ \mathcal{E} } / \sqrt{N}$, which keeps the error $\mathcal{E}$ constant for growing $N$, but prolongs the necessary preparation time $\tau_{GHZ}$.
These considerations, however, assume a weak driving $\Omega$, whereas for strong driving the pumping rate becomes limited by power broadening. We therefore need to use a suitably low driving strength $\Omega \sim \gamma / \sqrt{N}$  which sets a limit on the preparation rate.
These considerations and parameter values can be turned into a rigorous upper bound on the preparation time \cite{SI}:
\begin{align}
\tau_{\rm GHZ} \propto N^{3/2} \left({\rm log} N\right) \left({\rm log}\frac{1}{\mathcal{E}}\right) / (g \sqrt{\mathcal{E}}).
\label{tauGHZ}
\end{align}
We can thus prepare a GHZ state at any desired fidelity $F_{\rm GHZ} = 1 - \mathcal{E}$ within a preparation time that has a low-order ($N^{3/2}$) polynomial scaling in the number of qubits, for a coupling $g$ independent of $N$.
If instead the total preparation time is restricted to $t<T_{max}$, then the preparation error of an $N$-particle GHZ state will necessarily obey ${\mathcal E}>N\gamma^2/g^2$ and ${\mathcal E}>e^{-T_{max}\Gamma_+}$, limiting the achievable fidelity to $F_{\rm GHZ}\lesssim1-N/(gT_{max})^2$ \cite{SI}.

\begin{figure}[t]
\centering
\includegraphics[width=\columnwidth]{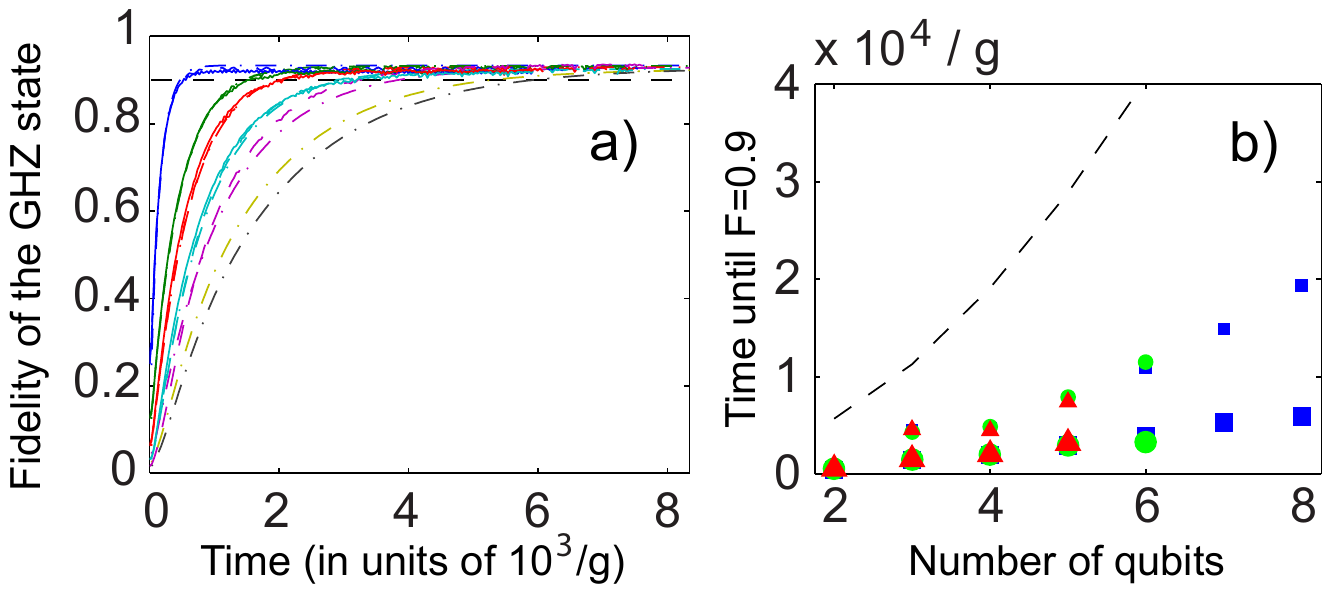}
\caption{Evolution towards a steady GHZ state.
Starting from a fully mixed state, we numerically solve an effective master equation. The curves in (a) show the evolution for two to eight qubits (different colors from blue to black) and are obtained by numerically optimizing all available parameters to reach a fidelity of $F_{\rm GHZ} = 0.9$ (black dashed line) within as short a time as possible. In (b) we show the scaling of the preparation time with the number of qubits. Both panels (a) and (b) show different degrees of truncation of the Hilbert space (dash-dots/blue squares -- effective ground state dynamics after adiabatic elimination, solid lines/green circles -- one excitation, dashes/red triangles -- two excitations). In (b), small symbols represent analytically optimized parameters and large symbols numerically optimized parameters. We find a polynomial scaling of the preparation time which is within our analytical bound (black dashed line in (b)).
}
\label{FigCurves}
\end{figure}

To confirm the results of the simple model we simulate the protocol numerically:
In Fig. \ref{FigCurves} (a) we plot the time-evolution towards a steady GHZ state for $N=2,\dots,8$ qubits resulting from our protocol. The plots are obtained by numerically simulating an effective master equation \cite{EO} as well as the complete master equation truncated to one or two excitations. Here we optimize the available parameters (driving strengths, tunable decay rates) to reach a fidelity $F_{\rm GHZ}=0.9$ of the GHZ state in minimum time. The resulting preparation times and the analytical bound from Eq. \eqref{tauGHZ} is shown in Fig. \ref{FigCurves} (b). These results confirm that the scheme exhibits a low-order polynomial scaling of the preparation time with the number of qubits. In contrast to the scheme in \cite{CBK}, our protocol requires only two operations for a GHZ state of $N$ qubits. Furthermore, the highly directed Z-pumping is only weakly perturbed by the polynomially weaker X-pumping and thus allows for a polynomial scaling of the protocol. 

The ingredients necessary for our scheme are available in trapped ion experiments. One suitable setup consists of a chain of $N$ trapped ions, each with two (meta-)stable ground levels $\ket{0}$ and $\ket{1}$ and two auxiliary levels, $\ket{e}$ and $\ket{f}$. Tunable decay of the auxiliary levels by spontaneous emission can be realized by a repumper to a higher lying rapidly decaying state \cite{Barreiro,Lin}. Two phononic modes, cooled to the ground state, and coupled to the sidebands of the ions' transitions, can be used as the harmonic oscillators $b$ and $c$. For the pumping, we require $2(N-1)$ field tones in the Z configuration and $2 \lfloor (N+1)/2 \rfloor$ tones in the X configuration. An alternative stroboscopic implementation requires only a single auxiliary level, interchanging between the roles of $\ket{e}$ and $\ket{f}$, a single phononic mode, interchanging between being $b$ and $c$, and a single field tone with tunable detuning applied on either the transition $\ket{1} \leftrightarrow \ket{e}$ or $\ket{-} \leftrightarrow \ket{f}$.
With $g/2\pi \sim 10$ kHz, typical preparation times are $\tau \sim 30$ ms. On such timescales collective dephasing needs to be taken into account, but can be overcome by switching the role of $\ket{0}$ and $\ket{1}$ in the Z-pumping for every second ion, thereby preparing $(\ket{0101...}+\ket{1010...})/\sqrt{2}$ which is in a decoherence free subspace and is equivalent to a GHZ state \cite{Monz}, or by using clock states \cite{Harty}. Fluctuations of $g$ of $1 \%$ result in a reduction of the fidelity by $0.01-0.1$ for $N=2, ..., 8$ qubits, whereas fluctuations of $0.1 \%$ have an effect at the sub-percent level. AC Stark shift fluctuations are suppressed since both red- and blue-detuned driving tones (e.g. $\Delta_\pm = \pm \sqrt{n_1} g$) are applied. Heating of the motion would constitute another error on the timescale of the scheme, but this can be made negligible for cryogenic traps \cite{Harty}.

We have shown that dissipative state preparation can be extended to the efficient generation of highly entangled steady states of many particles.
We achieve this by engineering complex multiparticle dissipation which deterministically drives the system into a desired steady state within a time scaling only polynomially with the size of the system. The generic couplings assumed in our approach can be found in a variety of physical systems, such as trapped ions where the basic ingredients of the scheme have already been demonstrated \cite{Lin}.
As specific examples we have considered the preparation of highly entangled GHZ states, which are paradigmatic multiparticle entangled states, but the developed techniques are applicable to a range of other quantum information tasks. Particularly relevant further possibilities are the construction of quantum error correcting codes \cite{ShorErrorCorrection, Steane} or the observation of exotic phase transitions \cite{Diehl} induced by multiparticle dissipation.

The research leading to these results has received funding under the European Union's Seventh Framework Programme (FP/2007-2013) through the ERC Grant Agreement QIOS (Grant No. 306576) and through SIQS (Grant No. 600645). This work was also supported by the Villum Kann Rasmussen Foundation. F.R. acknowledges helpful conversations with Yiheng Lin and support from the Studienstiftung des deutschen Volkes. D.R. was supported by the Marie Curie Fellowship QUINTYL under Grant No. 298742.

\begingroup
\renewcommand{\addcontentsline}[3]{}
\renewcommand{\section}[2]{}

\endgroup

\onecolumngrid

\section*{}

\newpage

\onecolumngrid

\renewcommand{\theequation}{S\arabic{equation}}

\def\>{\rangle}
\def\<{\langle}

\setlength{\parindent}{0pt}

\section*{Supplemental Material ``Scalable dissipative preparation of many-body entanglement''}

In this Supplemental Material to the Letter ``Scalable dissipative preparation of many-body entanglement'' we present the dynamical model of the generic system under consideration (Section \ref{SecModel}), the coupling configurations for the described protocols (Section \ref{SecCouplings}) and the resulting effective dynamics of the system (Section \ref{SecEffective}). We discuss the engineering of the dissipative many-body interactions for GHZ state preparation (Section \ref{SecEngineeredGHZ}) and strong driving effects (Section \ref{sectionstrongdrivingeffects}). In Section \ref{SecAnalysis} we analyze the scaling of the protocol both for weak and for strong driving.

\tableofcontents

\clearpage

\section{Dynamical model of the system}
\label{SecModel}

The dynamics of the open system is modeled by a master equation of Lindblad form
\begin{align}
\dot{\rho} = \mathcal{L}(\rho) = -i \left[ H,\rho \right] + \sum_k L_k \rho L_k^\dagger - \frac{1}{2} \left(L_k^\dagger L_k \rho + \rho L_k^\dagger L_k\right).
\label{EqMaster}
\end{align}
Here, $\rho$ is the density matrix of the system and $\mathcal{L}$ denotes a Liouvillian of Lindblad form. The Hamiltonian $H$ contains the coherent interactions, while sources of dissipation present in the system are represented as Lindblad (`jump') operators $L_k$.
The Hamiltonian for the coupling configurations detailed in Fig. 2 b)-c) is, in its most general form, given by
\begin{align}
H = H_{\rm free} + H_{\rm int} + H_{\rm drive}.
\end{align}
We consider a free Hamiltonian $H_{\rm free}$ which contains the energies of the levels of the $N$ atoms and the harmonic oscillator modes,
\begin{align}
H_{\rm free} &= \omega_{e} J_{ee} + \omega_{f} J_{ff} + \omega_b b^\dagger b + \omega_c c^\dagger c.
\end{align}
Here we have introduced $J_{ij} = \sum_{a=1}^N \sigma^{ij}_a = \sum_{a=1}^N \ket{i}_a \bra{j}$ and made the simplifications $\omega_0 = \omega_1 = 0$ and $\hbar = 1$.
An interaction Hamiltonian describes the atom-oscillator coupling, and a drive Hamiltonian $H_{\rm drive}$ contains the fields used to perform coherent excitations of the system. The interaction terms and drives required for GHZ preparation are detailed in Section \ref{SecCouplings}.
\\

The excited degrees of freedom in the system are generally subject to dissipation. Here the excited states $\ket{e}$ and $\ket{f}$ of each atom undergo spontaneous emission to each of the ground states $\ket{0}$ and $\ket{1}$, described by the jump operators
\begin{align}
\label{LindbladFirst}
L_{\gamma_{0e}, a} &= \sqrt{\gamma_{0e}} \ket{0}_a \bra{e}
\\
L_{\gamma_{1e}, a} &= \sqrt{\gamma_{1e}} \ket{1}_a \bra{e}
\\
L_{\gamma_{0f}, a} &= \sqrt{\gamma_{0f}} \ket{0}_a \bra{f}
\\
L_{\gamma_{1f}, a} &= \sqrt{\gamma_{1f}} \ket{1}_a \bra{f},
\end{align}
where the subscript $a$ denotes the atom number. The total decay rates of the excited levels are given by $\gamma_e = \gamma_{0e} + \gamma_{1e}$ and $\gamma_f = \gamma_{0f} + \gamma_{1f}$. For simplicity we will assume equal decay rates to both ground states, which is, however, not crucial for the protocol. In addition, we assume decay of excitations of the two oscillator modes, $b$ and $c$, represented by
\begin{align}
L_{\kappa_b} &= \sqrt{\kappa_b} b
\\
L_{\kappa_c} &= \sqrt{\kappa_c} c.
\label{LindbladLast}
\end{align}
The scheme does not require oscillator decay at all ($\kappa_b = \kappa_c = 0$). It may, however, still be useful to avoid heating.

\section{Coupling configurations}
\label{SecCouplings}

The atom-oscillator couplings of the four coupling configurations in Fig. 2 b)-c) of the main text are described by the interaction Hamiltonian
\begin{align}
H_{\rm int} &= H_{{\rm int}, Z} + H_{{\rm int}, X}
\\
H_{{\rm int}, Z} &= g \left(a^\dagger J_{1e} + a J_{1e}^\dagger \right)
\label{HintZ}
\\
H_{{\rm int}, X} &= g \left(b^\dagger J_{- f} + b J_{- f}^\dagger \right)
\label{HintX}
\end{align}
Here, `Z' and `X' denote the coupling configurations introduced in the main text (recall that $\ket{-} = (\ket{0}-\ket{1})/\sqrt{2}$. These terms describe that an atomic excitation can be exchanged coherently with the respective harmonic oscillator with a coupling constant $g$. By $H_{{\rm int},Z}$ an atomic excitation in $\ket{e}$ is exchanged with the oscillator $b$, leaving the atom in $\ket{1}$. $H_{{\rm int}, X}$ couples the excited level $\ket{f}$ to $\ket{-} = \frac{1}{\sqrt{2}}(\ket{0} + \ket{1})$ while exchanging the atomic excitation with the oscillator $c$. 
The coherent excitation of the atoms by classical driving fields is modeled by the drive Hamiltonian
\begin{align}
H_{\rm drive} &= H_{{\rm drive}, Z} + H_{{\rm drive}, X}
\\
H_{{\rm drive}, Z} &= \frac{1}{2} \sum_F \Omega_{\rm Z}^{(F)} e^{-i \omega_{\rm Z}^{(F)} t} J_{e1} + H.c.
\label{HdriveZ}
\\
H_{{\rm drive}, X} &= \frac{1}{2} \sum_F \Omega_{\rm X}^{(F)} e^{-i \omega_{\rm X}^{(F)} t} J_{f -} + H.c.
\label{HdriveX}
\end{align}
Here, we generally allow for several field tones with Rabi frequencies $\Omega_k^{(F)}$ and frequencies $\omega_{k}^{(F)}$, from which we will deduce detunings $\Delta_k^{(F)}$ of the respective fields.

\section{Effective dynamics of the open system}
\label{SecEffective}

In the following, we provide a detailed analysis of the effective dynamics of the system described in Section \ref{SecModel}. To this end, we briefly introduce the effective formalism presented in Ref. \cite{EO} and use it to derive effective operators for the coupling configurations presented in Section \ref{SecCouplings}.

\subsection{Effective operator formalism}

As can be seen from Section \ref{SecModel}, the dissipation affects the excited levels $\ket{e}$ and $\ket{f}$ and the oscillator modes $a$ and $b$. For weak driving the decaying degrees of freedom can be adiabatically eliminated from the master equation. This is done using the effective operator formalism presented in Ref. \cite{EO}. In this way, the dynamics of the master equation are reduced to effective couplings between the ground states of the system, described by an effective master equation
\begin{align}
\dot{\rho} = \mathcal{L}_{\rm eff}(\rho) = -i \left[H_{\rm eff}, \rho\right] + \sum_{k} L_{k, \rm eff} \rho (L_{k, \rm eff})^{\dagger} - \frac{1}{2}\left((L_{k, \rm eff})^{\dagger} L_{k, \rm eff} \rho + \rho (L_{k, \rm eff})^{\dagger} L_{k, \rm eff}\right) .
\label{EqEffectiveMaster}
\end{align}
Since we are dealing with multiple field tones $F$ that give rise to the effective couplings, we use the extended formalism for many fields (cf. \cite{EO}) with
\begin{align}
H_{\rm eff} &= - \frac{1}{2} V_- \sum_{F} \left( H_{{\rm NH}}^{(F)} \right)^{-1} V_+^{(F)} + H.c.
\label{Heff}
\\
L_{k, \rm eff} &= L_k \sum_F \left( H_{{\rm NH}}^{(F)} \right)^{-1} V_+^{(F)},
\label{Leff}
\\
H_{{\rm NH}}^{(F)} &= H_{\rm free} + H_{\rm int} - \frac{i}{2} \sum_k L_k^\dagger L_k - \omega^{(F)}
\end{align}
Here, $V_+^{(F)}$ denotes the exciting part and $V_-^{(F)}$ the deexciting part of the drive $V_\pm = \sum_F V_\pm^{(F)}$. The non-Hermitian Hamiltonian $H_{\rm NH}^{(F)}$, which contains the frequency $\omega^{(F)}$, describes the time evolution of the excited states.
The drives are regarded as perturbations and thus denoted by ``$V$''; they are defined by the drive Hamiltonians $H_{\rm drive}$ in Section \ref{SecCouplings} so that we use $V = H_{\rm drive}$ when we derive the effective operators for each of the coupling configurations below. Based on the assumption of weak excitation we will restrict our discussion to the states which have at most one atomic or oscillator excitation. $H_{\rm NH}$ then contains the energies and couplings of the excited states. Since $H_{\rm NH}$ needs to be inverted to compute the effective operators, we will, for each coupling situation, start out by discussing this entity.

\subsection{Derivation of the effective operators for the coupling configurations}

\subsubsection{Z configuration}
\label{SecEffOpZ}
We begin with the Z coupling configuration, which is similar to the X configuration. Both are required for the generation of GHZ states. We use $H_{{\rm int}, Z}$, $H_{{\rm drive}, Z}$ from Eqs. \eqref{HintZ} and \eqref{HdriveZ} and the Lindblad operators in Eqs. \eqref{LindbladFirst}--\eqref{LindbladLast}
to set up the non-Hermitian Hamiltonian
\begin{align}
H_{{\rm NH},Z}^{(F)} &= \tilde{\Delta}_Z^{(F)} J_{ee} + \tilde{\delta}_Z^{(F)} b^\dagger b + g \left(b^\dagger J_{1e} + b J_{1e}^\dagger \right).
\end{align}
Here we have introduced complex detunings $\tilde{\Delta}_Z^{(F)} = \omega_e - \omega_Z^{(F)} - i \gamma_e/2$ and $\tilde{\delta}_Z^{(F)} = \omega_{b} - \omega_Z^{(F)} - i \kappa_b/2$, where $F$ denotes the particular tone of the driving field. Furthermore we have changed into a frame rotating with the frequency of the drive $\omega_{\rm Z}^{(F)}$. For the derivation of the effective operators we will for simplicity drop the sub- and superscripts denoting the coupling configurations and field tones.
To invert the non-Hermitian Hamiltonian we divide it into four blocks
\begin{align}
H_{\rm NH} &= A + B + C + D \\
A = \tilde{\delta} b^\dagger b, ~ B = &g b^\dagger J_{1e}, ~ C = g b J_{1e}^\dagger, ~ D = \tilde{\Delta} J_{ee}
\end{align}
After this separation we can formally invert the Hamiltonian $H_{\rm NH}$, using Banachiewicz' theorem \cite{Banachiewicz} for the blockwise inversion of a square matrix,
\begin{align}
H_{\rm NH}^{-1} = a + b + c + d
\\
d = \left(D - C A^{-1} B\right)^{-1}, ~ a = A^{-1} + A^{-1} B ~ d ~ C A^{-1}, ~ b &= - A^{-1} B ~ d, ~ c = B^T = - d ~ C A^{-1}
\end{align}
We now need to compute $d$ to obtain any of the above elements. As we shall see, it is possible to simplify the calculation and to obtain closed expressions for the decay rates if we separate the involved operators by the number of atoms in state $\ket{1}$. This is done by introducing projection operators $P_{n_1}$ which project on the states with the same number $n_1$ (from now on, $n$) of atoms in $\ket{1}$.
For the Z and X configurations discussed here, $n_1$ (or $n_-$ for X), is conserved under the couplings by the coherent interactions $H_{\rm int}$ and $V$, but can be changed by the dissipative jump processes $L_k$, e.g. from $n_1$ to $n_1-1$ in the case of Z.
Using these projectors we can split the non-Hermitian Hamiltonian of the excited states and its four blocks by $n$,
\begin{align}
H_{\rm NH} &= \sum_{n=0}^N H_{{\rm NH},n} P_n =  \sum_{n=0}^N A_{n} + B_{n} + C_{n} + D_{n}
\end{align}
The inverse of these non-Hermitian Hamiltonians for each $n$ are then found to be
\begin{align}
H_{{\rm NH},n}^{-1} &= a_{n} + b_{n} + c_{n} + d_{n}
\end{align}
The effective operators of Eqs. \eqref{Heff}--\eqref{Leff} are formally given by
\begin{align}
L_{\kappa,\rm eff}
&= \sum_{n} \sqrt{\kappa} ~ b ~ b_{n} V P_n
\\
L_{\gamma 0,a, \rm eff}
&= \sum_{n} \sqrt{\gamma_0} \ket{0}_a \bra{e} d_{n} V P_n
\\
L_{\gamma 1,a, \rm eff}
&= \sum_{n} \sqrt{\gamma_1} \ket{1}_a \bra{e} d_{n} V P_n
\\
H_{\rm eff}
&= - \frac{1}{2} V \sum_{n} d_{n} V P_n  + H.c.
\end{align}
To obtain the effective Lindblad operators and the effective Hamiltonian it is thus sufficient to compute the blocks $d_n$ and $b_n$ of the inverse non-Hermitian Hamiltonian. Using the identities $a (a^\dagger a)^{-1} a^\dagger = \mathbbm{1}$ and $ P_g J_{1e} J_{ee} = P_g J_{1e}$ (where $P_g$ is the projector onto the ground states) we obtain
\begin{align}
d_{n} &= \frac{1}{\tilde{\Delta}} \left[J_{ee} - \left(n - \frac{\tilde{\Delta} \tilde{\delta}}{g^2}\right)^{-1} J_{1e}^\dagger J_{1e} \right]
\\
b_{n} &= \frac{g}{\tilde{\Delta} \tilde{\delta}} a^\dagger J_{1e} \left[J_{ee} - \left(n - \frac{\tilde{\Delta} \tilde{\delta}}{g^2}\right)^{-1} J_{1e}^\dagger J_{1e} \right]
\end{align}
With this, and readopting the sub- and superscripts for the configuration and the field, and changing to a more detailed notation for the effective Lindblad operators we find
\begin{align}
L_{\kappa_b, {\rm Z}}
&= \sum_{n_1,F} \frac{\sqrt{\kappa_b} \Omega_Z^{(F)}}{2} e^{-i \omega_Z^{(F)} t} \left(g - \frac{\tilde{\Delta}_Z^{(F)} \tilde{\delta}_Z^{(F)}}{n_1 g}\right)^{-1} P_{n_1}
\equiv \sum_{n_1,F} \frac{\sqrt{\kappa_b} \Omega_Z^{(F)}}{2 \tilde{g}_{Z,n_1}^{(F)}} e^{-i \omega_Z^{(F)} t} P_{n_1}.
\label{effLZfields1}
\\
L_{\gamma_{0e}, a, {\rm Z}}
&= \sum_{n_1, F} \frac{\sqrt{\gamma_{0e}} \Omega_Z^{(F)}}{2} e^{-i \omega_Z^{(F)} t} \left(\tilde{\Delta}_Z^{(F)} - \frac{n_1 g^2}{\tilde{\delta}_Z^{(F)}} \right)^{-1} \ket{0}_a \bra{1} P_{n_1}
\equiv \sum_{n_1,F} \frac{\sqrt{\gamma_{0e}} \Omega_Z^{(F)}}{2 \tilde{\Delta}^{(F)}_{Z,n_1}} e^{-i \omega_Z^{(F)} t} \ket{0}_a \bra{1} P_{n_1}
\\
L_{\gamma_{1e}, a, {\rm Z}}
&= \sum_{n_1, F} \frac{\sqrt{\gamma_{1e}} \Omega_Z^{(F)}}{2} e^{-i \omega_Z^{(F)} t} \left(\tilde{\Delta}_Z^{(F)} - \frac{n_1 g^2}{\tilde{\delta}_Z^{(F)}} \right)^{-1} \ket{1}_a \bra{1} P_{n_1}
\equiv \sum_{n_1,F} \frac{\sqrt{\gamma_{1e}} \Omega_Z^{(F)}}{2 \tilde{\Delta}^{(F)}_{Z,n_1}} e^{-i \omega_Z^{(F)} t} \ket{1}_a \bra{1} P_{n_1}
\label{effLZfields3}
\end{align}
To obtain this we have used the identities $P_g J_{1e} J_{e1} P_g = P_g J_{11} P_g$, $J_{11} P_{n_1} = n_1 P_{n_1}$, and $J_{ee} J_{e1} P_g = J_{e1} P_g$. In the last expression we have also introduced the effective detunings $\tilde{\Delta}_{Z,n_1}$ and the effective couplings $\tilde{g}_{Z,n_1}$ with
\begin{align}
\tilde{\Delta}^{(F)}_{Z,n_1} &= \tilde{\Delta}_Z^{(F)} - \frac{n_1 g^2}{\tilde{\delta}_Z^{(F)}}
\label{effDeltaZ}
\\
\tilde{g}^{(F)}_{Z,n_1} &= g - \frac{\tilde{\Delta}_Z^{(F)}\tilde{\delta}_Z^{(F)}}{n_1 g}.
\label{effgZ}
\end{align}
As can be seen from Eqs. \eqref{effLZfields1}--\eqref{effLZfields3}, dealing with multiple frequencies in the drive leads a priori to a sum over terms for all fields in the effective Lindblad operators. However, as the frequencies of these fields are well-distinguishable, we separate the Lindblad operators by their driving field $F$. Given the quadratic appearance of the Lindblad operators in the master equation, we can also drop the exponential phase factors. For the effective Lindblad operators for the fields $F$ we then obtain
\begin{align}
\label{effLZ1}
L_{\kappa, {\rm Z}}^{(F)} 
&= \sum_{n_1=0}^N \frac{\sqrt{\kappa_b} \Omega_Z^{(F)}}{2 \tilde{g}_{Z,n_1}^{(F)}} P_{n_1}
\\
\label{effLZ2}
L_{\gamma 0, a, {\rm Z}}^{(F)}
&= \sum_{n_1=0}^N \frac{\sqrt{\gamma_{0e}} \Omega_Z^{(F)}}{2 \tilde{\Delta}_{Z,n_1}^{(F)}} \ket{0}_a \bra{1} P_{n_1}
\\
\label{effLZ3}
L_{\gamma 1, a, {\rm Z}}^{(F)}
&= \sum_{n_1=0}^N \frac{\sqrt{\gamma_{1e}} \Omega_Z^{(F)}}{2 \tilde{\Delta}_{Z,n_1}^{(F)}} \ket{1}_a \bra{1} P_{n_1}
.
\end{align}
We also define the corresponding effective decay rates
\begin{align}
\kappa_{Z,n_1}^{(F)} &= \frac{\kappa_b (\Omega_Z^{(F)})^2}{4 |\tilde{g}_{Z,n_1}^{(F)}|^2}
\label{kappaZeff}
\\
\gamma_{0,Z,n_1}^{(F)} &= \frac{\gamma_{0e} (\Omega_Z^{(F)})^2}{4 |\tilde{\Delta}_{Z,n_1}^{(F)}|^2}
\label{gammaZeff1}
\\
\gamma_{1,Z,n_1}^{(F)} &= \frac{\gamma_{1e} (\Omega_Z^{(F)})^2}{4 |\tilde{\Delta}_{Z,n_1}^{(F)}|^2}
\label{gammaZeff2}
\end{align}
The operators in Eqs. \eqref{effLZ1}--\eqref{effLZ3} are then the effective Lindblad operators for the Z configuration. As can be seen from the expressions in Eqs. \eqref{effDeltaZ}--\eqref{effgZ}, the effective detunings $\tilde{\Delta}_{Z,n_1}^{(F)}$ can be made very small by a suitable choice of the frequencies $\omega_Z^{(F)}$ of the fields $F$ which can be used to engineer the rates $\gamma_{0,Z,n_1}^{(F)}$ and $\gamma_{1,Z,n_1}^{(F)}$ of the effective decay processes. The engineering of the effective decay process to prepare GHZ states will be subject to Section \ref{SecEngineeredGHZ}.
\\
We now turn to the effective Hamiltonian. The effective Hamiltonian is computed from Eq. \eqref{Heff}. Other than for the effective Lindblad operators, introducing a multi-tone driving field results in cross terms between different fields, here denoted by $F$ and $G$,
\begin{align}
H_Z
&= - \frac{1}{8} \sum_{n_1=0}^N \sum_{F,G} n_1 \Omega_Z^{(F)} \Omega_Z^{(G)} \left(\tilde{\Delta}_{Z,n_1}^{(F)} - \frac{n_1 g^2}{\tilde{\delta}_{Z,n_1}^{(F)}} \right)^{-1} e^{-i (\omega_Z^{(F)} - \omega_Z^{(G)}) t} P_{n_1} + H.c.
\\
&= - \sum_{n_1=0}^N \sum_{F,G} {\rm Re} \left( \frac{n_1 \Omega_Z^{(F)} \Omega_Z^{(G)}}{4 \tilde{\Delta}_{Z,n_1}^{(F)}} e^{-i (\omega_Z^{(F)} - \omega_Z^{(G)}) t} \right) P_{n_1} P_g
\end{align}
Here, all terms $F \neq G$ have fast rotating exponential phase factors. Restricting the treatment to $F = G$ where these terms cancel, we obtain the main contribution
\begin{align}
H_Z \approx - \sum_{n_1 = 0}^N \sum_F {\rm Re} \left( \frac{n_1 (\Omega_Z^{(F)})^2}{4 \tilde{\Delta}_{Z,n_1}^{(F)}} \right) P_{n_1}
\equiv \sum_{n_1=0}^N \sum_F s_{Z, n_1}^{(F)} P_{n_1},
\label{effHZ}
\end{align}

We thus find that the main effective Hamiltonian processes are AC Stark shifts with a magnitude
\begin{align}
s_{{\rm Z}, n_1}^{(F)} = - {\rm Re} \left( \frac{n_1 (\Omega_Z^{(F)})^2}{4 \tilde{\Delta}_{Z,n_1}^{(F)}} \right)
\label{effShiftsZ}
\end{align}
As we will see further below, our choice of the field tones will make these Hamiltonian terms compensate each other.

\subsubsection{X configuration}
\label{SecEffOpX}
We perform an analogous treatment for the X pumping mediated by the excited level $\ket{f}$ and the oscillator mode $c$, using the couplings in Eqs. \eqref{HintX} and \eqref{HdriveX}. We start with the non-Hermitian Hamiltonian
\begin{align}
H_{\rm NH,X}^{(F)} &= \tilde{\Delta}_X^{(F)} J_{ff} + \tilde{\delta}_X^{(F)} c^\dagger c + g c^\dagger J_{- f} + g c J_{- f}^\dagger 
\end{align}
with the complex energies $\tilde{\Delta}_X^{(F)} = \omega_f - \omega_X^{(F)} - i \gamma_f / 2$ and $\tilde{\delta}_X^{(F)} = \omega_{c} - \omega_X^{(F)} - i \kappa_c/2$. Carrying out the derivation in the same manner as above for Z, we obtain for the effective Lindblad operators
\begin{align}
L_{\kappa, X}^{(F)} 
&= \sum_{n_-=0}^N \frac{\sqrt{\kappa_c} \Omega_{\rm X}^{(F)}}{2 \tilde{g}_{X,n_-}^{(F)}} P_{n_-}
\label{effLX1}
\\
L_{\gamma 0, a, X}^{(F)}
&= \sum_{n_-=0}^N \frac{\sqrt{\gamma_{0f}} \Omega_{\rm X}^{(F)}}{2 \tilde{\Delta}_{X,n_-}^{(F)}} \ket{0}_a \bra{-} P_{n_-}
\label{effLX2}
\\
L_{\gamma 1, a, X}^{(F)}
&= \sum_{n_-=0}^N \frac{\sqrt{\gamma_{1f}} \Omega_{\rm X}^{(F)}}{2 \tilde{\Delta}_{X,n_-}^{(F)}} \ket{1}_a \bra{-} P_{n_-}
\label{effLX3}
,
\end{align}
with the effective detunings
\begin{align}
\tilde{\Delta}^{(F)}_{X,n_-} &= \tilde{\Delta}_X^{(F)} - \frac{n_- g^2}{\tilde{\delta}_X^{(F)}}
\label{effDeltaX}
\\
\tilde{g}^{(F)}_{X,n_-} &= g - \frac{\tilde{\Delta}_X^{(F)}\tilde{\delta}_X^{(F)}}{n_- g}
\label{effgX}
\end{align}
The effective decay rates can be written as
\begin{align}
\kappa_{X,n_-}^{(F)} &= \frac{\kappa_c (\Omega_{\rm X}^{(F)})^2}{4 |\tilde{g}_{X,n_-}^{(F)}|^2}
\label{kappaXeff}
\\
\gamma_{0,X,n_-}^{(F)} &= \frac{\gamma_{0f} (\Omega_X^{(F)})^2}{4 |\tilde{\Delta}_{X,n_-}^{(F)}|^2}
\label{gammaXeff1}
\\
\gamma_{1,X,n_-}^{(F)} &= \frac{\gamma_{1f} (\Omega_X^{(F)})^2}{4 |\tilde{\Delta}_{X,n_-}^{(F)}|^2}
\label{gammaXeff2}
\end{align}
These operators resemble the ones for Z pumping in Eqs. \eqref{effLZ1}--\eqref{effLZ3} if $\ket{1}$ is replaced by $\ket{-}$. The only difference is that spontaneous emission is still assumed to lead to the final states $\ket{0}$ and $\ket{1}$.

\subsection{Reduction of the dynamics to rate equations}
\label{SecRate}

Using the effective operator concept, we have so far reduced the dynamics of the open system to an effective master equation of the ground states. This description is exact to second order perturbation theory. In addition to that, we will later achieve a compensation of the effective Hamiltonian, $H_{\rm eff} = 0$, so that the remaining dynamics are purely dissipative.
We can then, in another step, reduce the complexity of the dynamics to rate equations of the populations. This is achieved by choosing subspaces of interest between which the interactions present in the system do not build up coherences. These subspaces are defined by projection operators, e.g. $P_A$ and $P_B$.
For negligible coherences between the subspaces, we can then trace over the Liouvillian evolution to obtain decay rates from subspace $A$ to subspace $B$
\begin{align}
\Gamma_{A \rightarrow B} &\equiv {\rm Tr}(P_B \mathcal{L}(P_A \rho_i P_A) P_B) \approx \sum_k {\rm Tr} (P_B L_k P_A \rho_i P_A L_k^\dagger P_B)
\\
&= \sum_k \sum_f \bra{\psi_f} P_B L_k P_A \rho_i P_A L_k^\dagger P_B \ket{\psi_f}
\end{align}
For the subspaces we will consider, the decay rate is the same for all states in the subspace. We can then calculate the decay rates using a single state $\ket{\psi_i}$, 
\begin{align}
\Gamma_{A \rightarrow B, k}
&= \sum_f |\bra{\psi_f} P_B L_k P_A \ket{\psi_i}|^2
\label{PureRate}
\end{align}
We will use this rate equation approach to analyze the scaling of the preparation time of the protocol in Section \ref{SecAnalysis}.

\subsection{Numerical simulations}

The evolution of the system is simulated by numerically solving the effective master equation \eqref{EqEffectiveMaster}, including the effective operators derived above. For the evolution due to the effective master equation of the protocol we use a Trotter-like ansatz, simulating the evolution under the Z and X coupling in an interchanging manner, performing base transformations between the eigenbases of $\sigma_z$ and $\sigma_x$ in-between. To verify our findings, we compare our result to the solution of the master equation \eqref{EqMaster} after truncation to one or two excitations.

\section{Engineering dissipative mechanisms for GHZ state preparation}
\label{SecEngineeredGHZ}

In this section we show how the effective operators derived in the previous section are engineered to prepare GHZ states efficiently. As is discussed in the main manuscript, the engineered operators are used to empty certain states and transfer the population to others in such a way that in the end only the target state remains. This is achieved using the driving fields in the coupling configurations Z and X which activate the effective decay processes derived in Section \ref{SecEffective}. By choosing suitable detunings for these driving fields we can then engineer the rates of the effective decay processes. The choice of the Rabi frequencies of these fields is subject to Section \ref{SecAnalysis} where we optimize the preparation time of the protocols.

\subsection{Preparing $\ket{\rm GHZ}$: the Z pumping}

The first mechanism to prepare a $\ket{\rm GHZ}$ state described in the main manuscript is the Z pumping: This process transfers the population of all states other than $\ket{0}^{\otimes N}$ and $\ket{1}^{\otimes N}$ to $\ket{0}^{\otimes N} = \frac{1}{\sqrt{2}}(\ket{\rm GHZ} + \ket{\rm GHZ_-})$, and thereby, to $\ket{\rm GHZ}$, without affecting the GHZ state.
We engineer this process by applying drives $\Omega^{(F)}_{Z \pm}$ in the Z configuration with detunings $\Delta^{(F)}_{Z \pm} = \delta^{(F)}_{Z \pm} = \pm \sqrt{F} g$, where $1 \leq F \leq N-1$. Here, the subscript $Z+$ denotes the fields with positive detunings and $Z-$ those with negative detunings. If the field index $F$ coincides with the number of atoms in $\ket{1}$, $n_1$, for a certain initial state and driving field, the field is resonant and the effective detunings in Eqs. \eqref{effDeltaZ}--\eqref{effgZ} become
\begin{align}
\tilde{\Delta}_{Z \pm,n_1}^{(F=n_1)} 
&= \tilde{\Delta}^{(F)}_{Z \pm} - \frac{n_1 g^2}{\tilde{\delta}^{(F)}_{Z \pm}}
\approx - \frac{i}{2} (\gamma_e + \kappa_b)
\\
\tilde{g}_{Z \pm,n_1}^{(F=n_1)} 
&= g - \frac{\tilde{\Delta}^{(F)}_{Z \pm} \tilde{\delta}^{(F)}_{Z \pm}}{n_1 g}
\approx \frac{i}{2} \frac{\gamma_e + \kappa_b}{\sqrt{n_1}},
\end{align}
Since we generally work in the strong coupling limit $\gamma, \kappa \ll g$, the above effective detunings are small compared to those for off-resonant driving fields with $n_1 \neq F$,
\begin{align}
\tilde{\Delta}_{Z \pm,n_1}^{(F \neq n_1)}
&\approx \pm \frac{F - n_1}{\sqrt{F}} g
\\
\tilde{g}_{Z \pm,n_1}^{(F \neq n_1)} 
&\approx \frac{n_1 - F}{n_1} g,
\end{align}
With the effective detunings we can then compute the effective decay rates in Eq. \eqref{kappaZeff} -- \eqref{gammaZeff2} for the processes with $F = n_1$ and for the off-resonant ones with $F \neq n_1$,
\begin{align}
& \kappa_{Z \pm, n_1}^{(F = n_1)} = \frac{n_1 \kappa_b (\Omega_Z^{(F)})^2}{(\gamma_e + \kappa_b)^2},
& \kappa_{Z \pm, n_1}^{(F \neq n_1)} = \frac{n_1^2 \kappa_b (\Omega_Z^{(F)})^2}{4 (n_1 - F)^2 g^2}.
\\
& \gamma_{0, Z \pm, n_1}^{(F = n_1)}  = \frac{\gamma_{0e} (\Omega_Z^{(F)})^2}{(\gamma_e + \kappa_b)^2},
& \gamma_{0, Z \pm, n_1}^{(F \neq n_1)} = \frac{F \gamma_{0e} (\Omega_Z^{(F)})^2}{4 (F - n_1)^2 g^2},
\\
& \gamma_{1, Z \pm, n_1}^{(F = n_1)}  = \frac{\gamma_{1e} (\Omega_Z^{(F)})^2}{(\gamma_e + \kappa_b)^2},
& \gamma_{1, Z \pm, n_1}^{(F \neq n_1)} = \frac{F \gamma_{1e} (\Omega_Z^{(F)})^2}{4 (F - n_1)^2 g^2},
\end{align}
From these expressions we can clearly see that the first group of rates is engineered to be strong, while the second group of rates is engineered to be suppressed.
Using the quantities above we obtain for the effective Lindblad operators in Eqs. \eqref{effLZ1}--\eqref{effLZ3}.
\begin{align}
L_{\kappa,Z \pm}^{(F)}
&\approx \sum_{n_1=1}^{N-1} \left[ \sqrt{\kappa_{Z \pm, n_1}^{(F = n_1)}} + \mathcal{O}\left(\frac{1}{g^2}\right) \right] P_{n_1} + \sqrt{\kappa_{Z \pm, n_1}^{(F \neq N)}} P_{N}, ~ ~ ~ (1 \leq F \leq N -1)
\\
L_{\gamma 0,a,Z \pm}^{(F)}
&\approx \sum_{n_1 = 1}^{N-1} \left[\sqrt{\gamma_{0, Z \pm, n_1}^{(F = n_1)}} \ket{0}_a \bra{1} + \mathcal{O}\left(\frac{1}{g^2}\right)\right] P_{n_1} + \sqrt{\gamma_{0, Z \pm, n_1}^{(F \neq N)}} \ket{0}_a \bra{1} P_{N}, ~ ~ ~ (1 \leq F \leq N -1)
\\
L_{\gamma 1,a,Z \pm}^{(F)}
&\approx \sum_{n_1 = 1}^{N-1} \left[\sqrt{\gamma_{1, Z \pm, n_1}^{(F = n_1)}} \ket{1}_a \bra{1} + \mathcal{O}\left(\frac{1}{g^2}\right)\right] P_{n_1} + \sqrt{\gamma_{1, Z \pm, n_1}^{(F \neq N)}} \ket{1}_a \bra{1} P_{N}, ~ ~ ~ (1 \leq F \leq N -1)
\end{align}
From these expressions we see that for $1 \leq n_1 \leq N-1$ we will always find terms with $n_1 = F$ to zeroth order in $g^{-1}$, which are much larger rates than the ones with $n_1 \neq F$, which are to second order in $g^{-1}$. We can therefore drop the latter for $1 \leq n_1 \leq N-1$. The terms with $n_1 = N$ which in particular affect the GHZ state need, however, to be kept. Since the effective detunings are engineered to depend on $n_1$, photons scattered by resonances with different $n_1$ can be distinguished. Formally this is justified by the exponential factors $e^{-i \omega t}$, washing out interferences between terms (cf. Section \ref{SecEffective}). We can therefore separate the terms with different $n_1$ into individual Lindblad operators, each acting on a set of states with $n_1$ atoms in $\ket{1}$. The enhanced processes are then given by
\begin{align}
L_{\kappa,Z \pm}^{(F=n_1)}
&\approx \sqrt{\kappa_{Z \pm,n_1}^{(F = n_1)}} P_{n_1=F}, ~ ~ ~ (1 \leq F \leq N -1)
\\
L_{\gamma 0,a,Z \pm}^{(F = n_1)}
&\approx \sqrt{\gamma_{0,Z \pm,n_1}^{(F = n_1)}} \ket{0}_a \bra{1} P_{n_1=F}, ~ ~ ~ (1 \leq F \leq N -1)
\\
L_{\gamma 1,a,Z \pm}^{(F = n_1)}
&\approx \sqrt{\gamma_{1,Z \pm,n_1}^{(F = n_1)}} \ket{1}_a \bra{1} P_{n_1=F}, ~ ~ ~ (1 \leq F \leq N -1)
\end{align}
The weak decay processes affecting $\ket{1}^{\otimes N}$, and thus, $\ket{\rm GHZ}$ are found to be
\begin{align}
L_{\kappa, Z \pm}^{(F \neq n_1)} \ket{\rm GHZ}
&\approx \sqrt{\kappa_{Z \pm, n_1}^{(F \neq n_1)}} P_{n_1=N} \ket{\rm GHZ}, ~ ~ ~ (1 \leq F \leq N -1) \\
&\approx \frac{1}{2} \sqrt{\kappa_{Z \pm, n_1}^{(F \neq n_1)}} (\ket{\rm GHZ} + \ket{\rm GHZ_-}) , ~ ~ ~ (1 \leq F \leq N -1)
\\
L_{\gamma 0,a, Z \pm}^{(F \neq n_1)} \ket{\rm GHZ}
&\approx \sqrt{\gamma_{0, Z \pm, n_1}^{(F \neq n_1)}} \ket{0}_a \bra{1} P_{n_1=N} P_{\rm GHZ}, ~ ~ ~ (1 \leq F \leq N - 1) \\
&\approx \frac{1}{2} \sqrt{\gamma_{0, Z \pm, n_1}^{(F \neq n_1)}} \ket{0}_a \bra{1} (\ket{\rm GHZ} + \ket{\rm GHZ_-}), ~ ~ ~ (1 \leq F \leq N - 1)
\\
L_{\gamma 1,a, Z \pm}^{(F \neq n_1)} \ket{\rm GHZ}
&\approx \sqrt{\gamma_{1, Z \pm, n_1}^{(F \neq n_1)}} \ket{1}_a \bra{1} P_{n_1=N} P_{\rm GHZ}, ~ ~ ~ (1 \leq F \leq N - 1) \\
&\approx \frac{1}{2} \sqrt{\gamma_{1, Z \pm, n_1}^{(F \neq n_1)}} \ket{1}_a \bra{1} (\ket{\rm GHZ} + \ket{\rm GHZ_-}), ~ ~ ~ (1 \leq F \leq N - 1)
\end{align}
The result of the engineering of the effective processes for the Z configuration is thus the Z pumping by which all population from states with $1 \leq n_1 \leq N-1$ is transferred to $\ket{0}^{\otimes N}$, and thus, to $\ket{\rm GHZ}$ and $\ket{\rm GHZ_-}$.

Beside effective decay processes, we generally also obtain effective Hamiltonian terms as given by Eq. \eqref{effHZ}, which we write as $H_Z = H_{Z+} + H_{Z-}$ with
\begin{align}
H_{Z \pm} \approx \mp \sum_{n_1 = 0}^N \frac{n_1}{4 g} \sum_{F \neq n_1} \frac{\sqrt{F} (\Omega_{\rm Z}^{(F)})^2}{F - n_1} P_{n_1}
\equiv \sum_{n_1=0}^N \sum_{F \neq n_1} s_{Z \pm, n_1}^{(F)} P_{n_1}.
\end{align}
These terms are AC Stark shifts, the magnitude of which is obtained from Eq. \eqref{effShiftsZ},
\begin{align}
s_{Z \pm, n_1}^{(F)} \approx \mp \frac{n_1 \sqrt{F} (\Omega_{\rm Z}^{(F)})^2}{4 (F - n_1) g}
\end{align}
By our combination of red-detuned drives ($\Omega^{(F)}_{Z+}$, $\Delta^{(F)}_{Z+}$) and blue-detuned drives ($\Omega^{(F)}_{Z-}$, $\Delta^{(F)}_{Z-}$) with $\Omega^{(F)}_{Z+} = \Omega^{(F)}_{Z-}$ and $\Delta^{(F)}_{Z +} = \Delta^{(F)}_{Z -}$ we achieve that the shifts compensate each other,
\begin{align}
H_Z \approx H_{Z+} + H_{Z-} \approx 0.
\label{effHcompensated}
\end{align}
Given that there are no Hamiltonian terms, we can turn to a description of the dynamics in terms of rate equations:

In Section \ref{SecRate}, we have described the possibility to reduce the effective dynamics further to rate equations. To perform such a step, we identify subspaces inside which all states have the same decay rate and in-between which no significant correlations are built up. It is therefore important that the effective Hamiltonian is zero.
We define the subspaces by the projectors $P_{\rm GHZ} = \ket{\rm GHZ}\bra{\rm GHZ}$, $P_{\rm GHZ_-} = \ket{\rm GHZ_-}\bra{\rm GHZ_-}$, and $P_{n_1}$ for $1 \leq n_1 \leq N-1$, which contain the states with a certain number of atoms in $\ket{1}$. The decay rates can then be calculated using Eq. \eqref{PureRate}. For resonant Z pumping ($F = n_1$) from a subspace with $n_1$ atoms in $\ket{1}$ to one with $n_1 - 1$ due to spontaneous emission to $\ket{0}$ we obtain the rate
\begin{align}
\Gamma_{n_1 \rightarrow n_1 - 1, \gamma 0, Z \pm}^{(F=n_1)} \approx \sum_a \sum_k |\bra{\psi_k} P_{n_1-1} L_{\gamma 0, a, Z \pm}^{(F=n_1)} P_{n_1} \ket{\psi_j} |^2
\approx
\frac{n_1 \gamma_{0e} (\Omega_Z^{(F=n_1)})^2}{(\gamma_e + \kappa_b)^2}.
\label{gamma0processesforZpm}
\end{align}
The subscripts in the above and the following decay rates specify the initial subspace, the final subspace, the physical process, i.e. oscillator decay ($\kappa$), spontaneous emission to $\ket{0}$ ($\gamma_0$) or $\ket{1}$ ($\gamma_1$), and the pumping process (here: $Z$). As opposed to the resonant rate ($F = n_1$) above, the decay rates due to off-resonant fields can be written as
\begin{align}
\Gamma_{n_1 \rightarrow n_1 - 1, \gamma 0, Z \pm}^{(F \neq n_1)} &= \sum_F \sum_a \sum_{k \neq j} |\bra{\psi_k} P_{n_1-1} L_{\gamma 0, a, Z \pm}^{(F \neq n_1)} P_{n_1} \ket{\psi_j} |^2
\approx \frac{n_1 \gamma_{0e} }{4 g^2} \sum_F F \left(\frac{\Omega_Z^{(F)}}{F - n_1} \right)^2,
\\
\Gamma_{n_1 \rightarrow n_1, \gamma 1, Z \pm}^{(F \neq n_1)} &= \sum_F \sum_a \sum_{k \neq j} |\bra{\psi_k} P_{n_1} L_{\gamma 1, a, Z \pm}^{(F \neq n_1)} P_{n_1} \ket{\psi_j} |^2
\approx \frac{n_1 \gamma_{1e} }{4 g^2} \sum_F F \left(\frac{\Omega_Z^{(F)}}{F - n_1} \right)^2.
\end{align}
From these expressions follow the loss rates from $\ket{\rm GHZ}$ due to Z pumping:
\begin{align}
\Gamma_{{\rm GHZ} \rightarrow {\rm GHZ}_-, \kappa, Z \pm}^{(F \neq n_1)} &= \sum_F |\bra{{\rm GHZ}_-} L_{\kappa, Z \pm}^{(F \neq n_1)} \ket{{\rm GHZ}} |^2
\approx \frac{N^2 \kappa_b}{16 g^2} \sum_{F=1}^{N-1} \left(\frac{\Omega_{Z}^{(F)}}{N - F}\right)^2,
\\
\Gamma_{{\rm GHZ} \rightarrow N-1, \gamma 0, Z \pm}^{(F \neq n_1)} &= \sum_F \sum_{a=1}^{N} \sum_{k \neq j} |\bra{\psi_k} L_{\gamma 0, a, Z \pm}^{(F \neq n_1)} \ket{{\rm GHZ}} |^2
\approx \frac{N \gamma_{0e}}{8 g^2} \sum_{F=1}^{N-1} F \left(\frac{\Omega_{Z}^{(F)}}{N - F}\right)^2,
\\
\Gamma_{{\rm GHZ} \rightarrow {\rm GHZ}_-, \gamma 1, Z \pm}^{(F \neq n_1)} &= \sum_{F=1}^{N-1} \sum_{a=1}^{N} |\bra{{\rm GHZ}_-} L_{\gamma 1, a, Z \pm}^{(F \neq n_1)} \ket{{\rm GHZ}} |^2
\approx \frac{N \gamma_{1e}}{16 g^2} \sum_{F=1}^{N-1} F \left(\frac{\Omega_{Z}^{(F)}}{N - F}\right)^2.
\end{align}
Here we have neglected the gain of population in $\ket{\rm GHZ}$ from $\ket{\rm GHZ_-}$. Furthermore, we note that due to its scaling with $N^2$, loss from $\ket{\rm GHZ}$ by oscillator decay should be avoided. In addition, effective oscillator decay is not useful for the Z pumping process. As we will see below, this is also the case for X pumping. Therefore, we will generally choose to work with $\kappa_{b/c} = 0$ for the GHZ protocol (a weak cooling $\kappa \ll \Omega$ may nevertheless be used to counteract heating). 
The overall loss rate from $\ket{\rm GHZ}$ due to off-resonant Z pumping is then given by
\begin{align}
\Gamma_{{\rm GHZ} \rightarrow ?, Z} &\approx \frac{N (2\gamma_{0e} + \gamma_{1e})}{8 g^2} \sum_{F=1}^{N-1} F \left(\frac{\Omega_Z^{(F)}}{N - F}\right)^2.
\end{align}
Here, the question mark stands for any potential final state; in the scaling analysis we will typically consider the worst state possible. For the reasonable assumption of $\gamma_{0e} = \gamma_{1e} = \gamma_e / 2$ we obtain
\begin{align}
\Gamma_{{\rm GHZ} \rightarrow ?, Z} &\approx \frac{3 \gamma_e N}{16 g^2} \sum_{F=1}^{N-1} F \left(\frac{\Omega_Z^{(F)}}{N - F}\right)^2.
\label{gammaminusformula}
\end{align}
We conclude that for $g \gg \gamma, \kappa$, the rates from Z pumping, ultimately leading to $\ket{0}^{\otimes N}$, and thus to $\ket{\rm GHZ}$, are much stronger than the loss rates from $\ket{\rm GHZ}$. 
\\
The derived rates will be used further to analyze the error and preparation time of the protocol in Section \ref{SecAnalysis}.

\subsection{Emptying $\ket{\rm GHZ_-}$: the X configuration}
\label{subsectiondepumpingGHZX}

$\ket{\rm GHZ_-}$ is emptied by X pumping, as described in the main part:
Here we make use of the fact that in the eigenbasis of $\sigma_x = (\ket{1}\bra{0}+\ket{0}\bra{1})$, consisting of $\ket{\pm} = (\ket{0} \pm \ket{1})/\sqrt{2}$, $\ket{\rm GHZ_-}$ is a superposition of states with odd numbers of qubits in $\ket{-}$, whereas $\ket{\rm GHZ}$ only contains states with an even number of qubits in $\ket{-}$,
\begin{align}
\ket{\rm GHZ_-} &= \frac{1}{\sqrt{2}^{N-1}}\left(\left(\ket{+...++-} + \ket{+...+-+} + ...\right) + ...\right).
\\
\ket{\rm GHZ} &= \frac{1}{\sqrt{2}^{N-1}}\left(\ket{+...+++} + \left(\ket{+...+--} + \ket{+...--+} + ...\right) + ...\right),
\label{GHZX}
\end{align}
To depump $\ket{\rm GHZ_-}$ without affecting $\ket{\rm GHZ}$ we apply again red- and blue-detuned laser fields $\Omega^{(F)}_{X \pm}$ with detunings $\Delta_{X \pm}^{(F)} = \pm \sqrt{F} g$, but including only odd field indices $F = 1, 3, 5, ..., (F \leq N)$, whereas for even $F = 2, 4, ...$ we use $\Omega_X^{(F)} = 0$. From Eqs. \eqref{effDeltaX}--\eqref{effgX} we find for the effective detunings for resonant ($F = n_-$) and off-resonant ($F \neq n_-$) excitation
\begin{align}
&\tilde{\Delta}_{X \pm, n_-}^{(F = n_-)} 
= \tilde{\Delta}^{(F)}_{X \pm} - \frac{n_- g^2}{\tilde{\delta}^{(F)}_{X \pm}}
\approx - \frac{i}{2} (\gamma_f + \kappa_c),
&\tilde{\Delta}_{X \pm, n_-}^{(F \neq n_-)} 
\approx \pm \frac{F - n_-}{\sqrt{F}} g, \\
&\tilde{g}_{X \pm, n_-}^{(F = n_-)} 
= g - \frac{\tilde{\Delta}^{(F)}_{X \pm} \tilde{\delta}^{(F)}_{X \pm}}{n_- g}
\approx \frac{i}{2} \frac{\gamma_f + \kappa_c}{\sqrt{n_-}},
&\tilde{g}_{X \pm, n_-}^{(F \neq n_-)} 
\approx \frac{n_- - F}{n_-} g.
\end{align}
For the effective decay rates we find from Eqs. \eqref{kappaXeff}--\eqref{gammaXeff2},
\begin{align}
&\kappa_{X \pm, n_-}^{(F = n_-)} = \frac{n_- \kappa_c (\Omega_X^{(F)})^2}{(\gamma_f + \kappa_c)^2},
&\kappa_{X, n_-}^{(F \neq n_-)} = \frac{\kappa_c n_-^2 (\Omega_{X}^{(F)})^2}{4 (n_- - F)^2 g^2}.
\\
&\gamma_{0, X \pm, n_-}^{(F = n_-)}  = \frac{\gamma_{0f} (\Omega_X^{(F)})^2}{(\gamma_f + \kappa_c)^2},
&\gamma_{0, X \pm, n_-}^{(F \neq n_-)} = \frac{\gamma_{0f} F (\Omega_X^{(F)})^2}{4 (F - n_-)^2 g^2},
\\
&\gamma_{1, X \pm, n_-}^{(F = n_-)}  = \frac{\gamma_{1f} (\Omega_X^{(F)})^2}{(\gamma_f + \kappa_c)^2},
&\gamma_{1, X \pm, n_-}^{(F \neq n_-)} = \frac{\gamma_{1f} F (\Omega_X^{(F)})^2}{4 (F - n_-)^2 g^2},
\end{align}
for resonant excitation ($F = n_-$) and for off-resonant ($F \neq n_-$) excitation. With these rates and Eqs. \eqref{effLX1}--\eqref{effLX3} we obtain the effective Lindblad operators
\begin{align}
L_{\kappa, X \pm}^{(F)}
= &\sum_{{\rm odd} ~ n_-} \left[\sqrt{\kappa_{X \pm, n_-}^{(F = n_-)}} + \mathcal{O}\left(\frac{1}{g^2}\right)\right] P_{n_-}
+ \sum_{{\rm even} ~ n_-} \sqrt{\kappa_{X \pm, n_-}^{(F \neq n_-)}} P_{n_-}, ~ ~ ~ (F = 1, 3, 5, ..., (F \leq N)),
\\
L_{\gamma 0,a,X \pm}^{(F)}
= &\sum_{{\rm odd} ~ n_-} \left[\sqrt{\gamma_{0, X \pm, n_-}^{(F = n_-)}} \ket{0}_a \bra{-} P_{n_-} + \mathcal{O}\left(\frac{1}{g^2}\right)\right]
+ \sum_{{\rm even} ~ n_-} \sqrt{\gamma_{0, X \pm, n_-}^{(F \neq n_-)}} \ket{0}_a \bra{-} P_{n_-}, ~ ~ ~ (F = 1, 3, 5, ..., (F \leq N)),
\\
L_{\gamma 1,a,X \pm}^{(F)}
= &\sum_{{\rm odd} ~ n_-} \left[\sqrt{\gamma_{1, X \pm, n_-}^{(F = n_-)}} \ket{1}_a \bra{-} P_{n_-} + \mathcal{O}\left(\frac{1}{g^2}\right)\right]
+ \sum_{{\rm even} ~ n_-} \sqrt{\gamma_{1, X \pm, n_-}^{(n_- \neq F)}} \ket{1}_a \bra{-} P_{n_-}, ~ ~ ~ (F = 1, 3, 5, ..., (F \leq N)).
\end{align}
Again, we separate the effective Lindblad operators by the frequencies of the resonances, this time depending on $n_-$. We obtain similar Lindblad operators as for Z pumping, with enhanced terms to zeroth order in $g^{-1}$,
\begin{align}
L_{\kappa,X \pm}^{(F= n_-)}
&= \sqrt{\kappa_{X \pm, n_-}^{(F = n_-)}} P_{n_-}, ~ ~ ~ (F = 1, 3, 5, ..., (F \leq N)),
\label{effLXpump1}
\\
L_{\gamma 0,a,X \pm}^{(F = n_-)}
&= \sqrt{\gamma_{0, X \pm, n_-}^{(F = n_-)}} \ket{0}_a \bra{-} P_{n_-}, ~ ~ ~ (F = 1, 3, 5, ..., (F \leq N)),
\label{effLXpump2}
\\
L_{\gamma 1,a,X \pm}^{(F = n_-)}
&= \sqrt{\gamma_{1, X \pm, n_-}^{(F = n_-)}} \ket{1}_a \bra{-} P_{n_-}, ~ ~ ~ (F = 1, 3, 5, ..., (F \leq N)),
\label{effLXpump3}
\end{align}
and suppressed terms acting on the target state
\begin{align}
L_{\kappa,X \pm}^{(F \neq n_-)}
&= \sqrt{\kappa_{X \pm,n_-}^{(F \neq n_-)}} P_{n_-}, ~ ~ ~ ({\rm even} ~ n_-, {\rm odd} ~ F),
\label{effLXloss1}
\\
L_{\gamma 0,a,X \pm}^{(F \neq n_-)}
&= \sqrt{\gamma_{0,X \pm,n_-}^{(F \neq n_-)}} \ket{0}_a \bra{-} P_{n_-}, ~ ~ ~ ({\rm even} ~ n_-, {\rm odd} ~ F),
\label{effLXloss2}
\\
L_{\gamma 1,a,X \pm}^{(F \neq n_-)}
&= \sqrt{\gamma_{1,X \pm,n_-}^{(F \neq n_-)}} \ket{1}_a \bra{-} P_{n_-}, ~ ~ ~ ({\rm even} ~ n_-, {\rm odd} ~ F).
\label{effLXloss3}
\end{align}
It can be seen that for $g \gg \gamma, \kappa$ these operators make $\ket{\rm GHZ_-}$ (with only odd $n_-$) decay rapidly. The losses from $\ket{\rm GHZ}$ (with only even $n_-$) are caused by the off-resonant drives with $F \neq n_-$. Since these terms are of second order in $g^{-1}$, they are suppressed for $g \gg \gamma, \kappa$.

Again, the effective Hamiltonian is compensated by the red- and blue-detuned fields,
\begin{align}
H_{X} \approx H_{X+} + H_{X-} \approx 0.
\end{align}
We can therefore describe the dynamics in terms of rates:

With the effective operators from Eqs. \eqref{effLXpump1}--\eqref{effLXpump3} we find the decay rates from $\ket{\rm GHZ_-}$,
\begin{align}
\Gamma_{{\rm GHZ_-} \rightarrow ?, \gamma 0, X \pm}
&\approx \frac{\gamma_{0f}}{(\gamma_f + \kappa_c)^2} \sum_{F=1,3,...}^N {N \choose F} \frac{F (\Omega_{X}^{(F)})^2}{2^{N-1}},
\\
\Gamma_{{\rm GHZ_-} \rightarrow ?, \gamma 1, X \pm}
&\approx \frac{\gamma_{1f}}{(\gamma_f + \kappa_c)^2} \sum_{F=1,3,...}^N {N \choose F} \frac{F (\Omega_{X}^{(F)})^2}{2^{N-1}}.
\end{align}
For the reasonable assumption of $\gamma_{0f} = \gamma_{1f} = \gamma_f/2$ the total rate is therefore given by
\begin{align}
\Gamma_{{\rm GHZ_-} \rightarrow ?, X}
&\approx \frac{2 \gamma_f}{(\gamma_f + \kappa_c)^2} \sum_{F=1,3,...}^N {N \choose F} \frac{F (\Omega_{X}^{(F)})^2}{2^{N-1}}.
\label{GammaXtotal}
\end{align}
The expression in Eq. \eqref{GammaXtotal} is the total decay rate from the $\ket{\rm GHZ_-}$ state which is approximately given by the sum of all enhanced decay rates, weighted with the number of states with the same excitation. The X pumping distributes the population of $\ket{\rm GHZ_-}$ over all other states which we here denote by a question mark and later on replace by a worst-case assumption.
\\
The losses from $\ket{\rm GHZ}$ are only caused by the off-resonant drives with $F \neq n_-$.
Using Eqs. \eqref{effLXloss1}--\eqref{effLXloss3} and ignoring negligible gain processes we obtain the loss rates
\begin{align}
\Gamma_{{\rm GHZ} \rightarrow ?, \gamma 0, X \pm}
& \approx \frac{\gamma_{0f}}{4 g^2} \sum_{n_-=0,2,...} {N \choose n_-} \frac{n_-}{2^{N-1}} \sum_{F=1,3,...} F \left(\frac{\Omega_X^{(F)}}{F - n_-}\right)^2
\\
\Gamma_{{\rm GHZ} \rightarrow ?, \gamma 1, X \pm}
& \approx \frac{\gamma_{1f}}{4 g^2} \sum_{n_-=0,2,...} {N \choose n_-} \frac{n_-}{2^{N-1}} \sum_{F=1,3,...} F \left(\frac{\Omega_X^{(F)}}{F - n_-}\right)^2
\end{align}
Here, the binomial coefficients originate from the number of states with the same number of atoms in $\ket{-}$. For the total loss rate from $\ket{\rm GHZ}$ through X pumping with both red- ($X+$) and blue-detuned ($X-$) fields we approximately find 
\begin{align}
\Gamma_{{\rm GHZ} \rightarrow ?, \gamma, X}
\approx \frac{\gamma_{f}}{2 g^2} \sum_{n_-=0,2,...} {N \choose n_-} \frac{n_-}{2^{N-1}} \sum_{F=1,3,...} F \left(\frac{\Omega_X^{(F)}}{F - n_-}\right)^2
\label{GammaXminus}
\end{align}
From this expression we see that the loss terms from $\ket{\rm GHZ}$ due to X pumping are of second order in $g^{-1}$ and thus suppressed for $g \gg \gamma, \kappa$. Beside decay out of $\ket{\rm GHZ_-}$ the X pumping also causes losses from states with odd $n_-$ which have an overlap with states with $1 \leq n_1 \leq N-1$ in the Z basis. This affects the transport from $n_1 = N-1$ to $n_1 = 0$ by the Z pumping and thus to $\ket{\rm GHZ}$ by imposing a loss rate
\begin{align}
\Gamma_{n_1 \rightarrow ?, \gamma, X}
\approx \frac{\gamma_{f}}{(\gamma_f + \kappa_c)^2} \sum_{F=1,3,...}^N {N \choose F} \frac{F (\Omega_X^{(F)})^2}{2^{N-1}}.
\label{Xtossrate}
\end{align}
We will refer to this process as ``X toss'' below. As we will show below this does not, however, have a significant effect on the scaling of the preparation time and the error, if the strength of the X process is chosen properly.
We conclude that the combination of Z and X pumping results in the preparation of a GHZ state from any initial state. For $g \gg \gamma, \kappa$ the gain rates are engineered to be strong, while the loss rates from $\ket{\rm GHZ}$ are suppressed. The scaling of the error and the preparation time of the protocol are analyzed in Section \ref{SecAnalysis}.

\section{Strong driving effects}\label{sectionstrongdrivingeffects}

The operator formalism used to derive effective couplings of the ground states in the system is built on the assumption of a perturbative drive which is much smaller than the couplings, decay, or detunings of the excited states, $\Omega \ll \gamma, \kappa, g, \delta, \Delta$. Effects from saturating the excited states that could possibly slow down the preparation process are thus not included in the rates derived so far. On the contrary, the dependency of all decay rates on the drives $\propto \Omega^2$ suggests that the optimal Rabi frequencies $\Omega$ of the drives are infinitely strong and thereby outside the perturbative regime. A proper assessment of the driving strength $\Omega$ therefore requires the inclusion of strong driving effects which begin to play a role for $\Omega \lesssim \gamma, \kappa$.

In the following, we include such effects in the dynamics in an approximate way. Here, we discuss power broadening and population of the excited states and adjust the rates derived from the effective operators to account for power broadening of the excited states. They will later be used to analytically derive the optimal Rabi frequencies $\Omega$ and the scaling of the protocol for strong driving. In addition, in our numerical simulations we use effective operators where we have included the power broadening terms manually. While this treatment is not rigorous, it provides a convenient tool for rapidly determining the optimal parameters. These are used to simulate the effective dynamics of the system beyond the regime of weak driving and to match it with the simulations describing a larger Hilbert space including excitations. In addition, we take into account the effect of population of the excited states in our numerics.

\subsection{Power broadening}\label{powerbroadeningdescriptionsubsection}

We first address the effect of power broadening (or `line' broadening), regarding a simple model situation: A ground state $\ket{0}$ is resonantly coupled by a field with a Rabi frequency $\Omega$ to an excited level $\ket{e}$. In total, $\ket{e}$ decays at a rate $\gamma=\gamma_0+\gamma_1$, where $\gamma_1$ is the decay rate into $\ket{1}$. We perform adiabatic elimination by setting the derivative of the density matrix to zero, $\dot{\rho} \approx 0$. In the weak driving regime, where the broadening of the excited level can be neglected, this yields an excited population of $\rho_{\rm ee} = \Omega^2 / \gamma^2 \rho_{00}$ and thus an effective decay rate from $\ket{0}$ into $\ket{1}$ of $\gamma_{\rm eff} = \Omega^2/(2 \gamma)$. The population gain of state $\ket{1}$ is then given by
\begin{align}
\dot{\rho}_{11} &\approx \gamma_1 \rho_{ee} \approx \gamma_{\rm eff} \rho_{00} \approx \frac{\Omega^2}{2 \gamma} \rho_{00}
\end{align}
The same result is obtained when using the effective operators. Performing adiabatic elimination with a stronger drive we need to take into account the population of the excited level. This yields an excited population of $\rho_{ee} = \Omega^2 / (\gamma^2 + \Omega^2)$. Then, as the coupling of the ground state $\ket{0}$ to $\ket{\rm e}$ is increased, we need to take into account the population of the coupled subspace of $\ket{0}$ and $\ket{e}$ rather than of state $\ket{0}$ only. This leads to
\begin{align}
\dot{\rho}_{11} &\approx \frac{\gamma_1 \rho_{00}}{(\rho_{00} + \rho_{\rm ee})} (\rho_{00} + \rho_{\rm ee}) \approx \frac{\gamma_1 \Omega^2}{\gamma^2 + 2 \Omega^2} (\rho_{00} + \rho_{\rm ee})
\end{align}
In the weak driving limit $\Omega \rightarrow 0$, this new decay rate $\gamma_{\rm eff} = \gamma \Omega^2 / (2(\gamma^2 + 2 \Omega^2))$ matches the previous result. The decay rate for strong driving can thus obtained from the rate which was derived from the effective operators by ``broadening'' the natural line width of the excited state, $\gamma^2 \rightarrow \gamma^2 + 2 \Omega^2$. In the limit of strong driving $\Omega \rightarrow \infty$, the population is found in $\ket{e}$ with a probability of $1/2$ such that here $\gamma_{\rm eff}$ approaches the constant value of $\gamma_{1}/2$. Therefore, the effective decay rate from $\ket{0}$ to $\ket{1}$ cannot be increased beyond half of the line width of the level mediating it.

More generally, we now seek to include power broadening in the effective operators. This is done by replacing the complex effective detunings, e.g. $\tilde{\Delta}_{Z,n_1}^{(F)}$, here generally denoted as $\tilde{\Delta}_{\rm eff}$ and $\tilde{g}_{\rm eff}$, by ``power broadened'' ones. The replacements are made such that the rates obtained from the effective operators agree with the ones derived by adiabatic elimination. To this end, we make the replacements $|\tilde{\Delta}_{\rm eff}|^2 \rightarrow |\tilde{\Delta}_{\rm eff}|^2 + n \Omega^2$ and $|\tilde{g}_{\rm eff}|^2 \rightarrow |\tilde{g}_{\rm eff}|^2 + n \Omega^2$, where $n$ is the number of atoms that can be excited by the drive. From our numerical simulations it turns out that the action of power broadening needs to be doubled in the effective operators for Z and X to achieve high accuracy between the evolution due to the effective master equation \eqref{EqEffectiveMaster} and the more complete master equation in Eq. \eqref{EqMaster}. This can be attributed to interference between the blue- and red-detuned drives. Indeed, considering coherent excitation of a bright state consisting of both the blue- and the red-shifted dressed state suggests an increase of the broadening by a factor of two and thus the replacements of the effective detunings $|\tilde{\Delta}_{\rm eff}|^2 \rightarrow |\tilde{\Delta}_{\rm eff}|^2 + 2 n \Omega^2$ and $|\tilde{g}_{\rm eff}|^2 \rightarrow |\tilde{g}_{\rm eff}|^2 + 2 n \Omega^2$.
While these replacements are less rigorous, they are supported by our numerical simulations:

It can be seen from Fig. 3 in the main manuscript that (1) the analytical scalings derived using the strong driving operators comprise upper bounds to the numerically obtained scalings and that (2) the effective operators for strong driving agree well with simulations of the more complete master equation.
\\

Following the reasoning above, the power broadened decay rate for the Z pumping from $n_1$ to $n_1 - 1$ is found to be
\begin{align}
\Gamma_{n_1 \rightarrow n_1 - 1, \gamma 0, Z}
\approx \frac{2 n_1 \gamma_{0e} (\Omega_Z^{(F=n_1)})^2}{(\gamma_e + \kappa_b)^2 + 2 n_1 (\Omega_Z^{(F=n_1)})^2}.
\label{GammaGoodPower}
\end{align}
For the X depumping of $\ket{\rm GHZ_-}$ we have
\begin{align}
\Gamma_{{\rm GHZ_-} \rightarrow ?, \gamma, X}
\approx \sum_{F=1,3,...}^N \frac{2 \gamma_f}{(\gamma_f + \kappa_c)^2 + 2 F (\Omega_X^{(F=n_-)})^2}  {N \choose F} \frac{F (\Omega_X^{(F=n_-)})^2}{2^{N-1}}
\label{goodXratepowerbroadening}
\end{align}
and for the rate of the X process when acting on other states with $n_1$ (the ``X toss'' rate)
\begin{align}
\Gamma_{n_1 \rightarrow ?, \gamma, X}
\approx \sum_{F=1,3,...}^N \frac{\gamma_{f}}{(\gamma_f + \kappa_c)^2 + 2 F (\Omega_X^{(F=n_-)})^2}  {N \choose F} \frac{n_- (\Omega_X^{(F=n_-)})^2}{2^{N-1}}.
\label{tossXratepowerboradeningasdfqwerasdf}
\end{align}

As a consequence of the inclusion of terms for power broadening, increasing the driving strengths in the desired processes also increases the effect of power broadening in the desired decay rates. In the off-resonant decay rates $\gamma_{\rm eff} \propto \gamma \Omega/g^2$ the effect of power broadening is, on the other hand, negligible (given that $\Omega^2 \ll g^2$). The detrimental rates thus still increase for a growing $\Omega$ while the desired rates saturate. This is the limiting factor for the drive $\Omega$ which we will use to derive the possible $\Omega$ further down.

\subsection{Population of the excited states}

An additional reduction in the fidelity for strong driving comes from the population of the excited induced by the drive. In the following, we investigate this effect:
\\
The GHZ state, through its contribution from $\ket{1}^{\otimes N}$, is coupled to an excited state $\ket{\rm \psi_e} = (\ket{1..1e} + \ket{1..e1} + ... + \ket{e..11})/\sqrt{N}$. Despite being off-resonance, all tones of the driving field couple to the transition from $\ket{\rm GHZ}$ to the dressed states of $\ket{\psi_e}$ and $\ket{1}^{\otimes N} \ket{1}$, with driving strengths of $\Omega_Z^{(F)} = \sqrt{N/2} \Omega_Z^{(F)}$ and detunings $\Delta_{Z \pm}^{(F)} = (\sqrt{N} \pm \sqrt{F}) g$. For each tone the excited population is then approximately given by
\begin{align}
P_{\rm excited, Z \pm}^{(F)} \approx \frac{N \Omega^{(F)}_Z}{8 (\sqrt{N} + \sqrt{F})^2 g^2} + \frac{N \Omega^{(F)}_Z}{8 (\sqrt{N} - \sqrt{F})^2 g^2}
\end{align}
We also consider the excited population caused by X pumping. This requires representing the GHZ state in the X basis as in Eq. \eqref{GHZX} and leads to an expression
\begin{align}
P_{\rm excited, X \pm}^{(F)} \approx \sum_{n_-=2,4,...}^N {N \choose n_-} \left( \frac{n_- \Omega^{(F)}_X}{4 (\sqrt{n_-} + \sqrt{F})^2 g^2} + \frac{n_- \Omega^{(F)}_X}{4 (\sqrt{n_-} - \sqrt{F})^2 g^2} \right) 
\end{align}
These expressions are included in our numerical simulations of the effective dynamics in the strong driving regime and provide another limitation on $\Omega$. Our analytical derivation of the scaling, on the other hand, involves upper bounds to certain error processes and turns out to take place in a parameter regime where the population of the excited states is not significant. Therefore, in the analytical considerations below we leave out its effect.

\section{Scaling analysis of the preparation time and the error of the protocol}
\label{SecAnalysis}

In the following, we provide an analytical study of the scaling of the GHZ scheme. We derive an expression for the preparation time and optimize it by the choice of the parameters. The analysis is performed both for weak driving, beginning in Section \ref{SecPreparation}, and for strong driving, in Section \ref{SecGHZStrong}.

\subsection{Optimization of the parameters for Z pumping alone}
\label{SecPreparation}

The different schemes presented in the main manuscript have in common, that the population of a nearly exponential number of states is pumped to subspaces with a polynomial number of states in a number of steps linear in the size of the system. For GHZ state preparation we have engineered strong decay processes from $N-1 \geq n_1 \geq 1$. The decay is achieved using the Z pumping which is only sensitive to $n_1$ and reduces this to $n_1 - 1$, finally leading to $n_1 = 0$.
In order to assess the performance of the schemes it is thus important to know the time for a concatenated process consisting of many consecutive Z pumping steps. The rate of each individual decay is given by (see Eq.\ (\ref{gamma0processesforZpm})):
\begin{align}
\Gamma_{n_1 \rightarrow n_1 - 1, \gamma 0, Z}^{(F=n_1)} = \sum_{a=1}^N \sum_k |\bra{\psi_k} P_{n_1-1} L_{\gamma 0, a, {\rm Z}}^{(F=n_1)} P_{n_1} \ket{\psi_j} |^2
\approx \frac{2 n_1 \gamma_{0e} (\Omega_Z^{(F=n_1)})^2}{(\gamma_e + \kappa_b)^2}\label{rateasdfasdfasdfabasdbasdbasb}
\end{align}
The average time for this decay to occur is given by the inverse decay rate,
\begin{align}
\tau_{n_1 \rightarrow n_1 - 1, \gamma 0, Z} = \Gamma_{n_1 \rightarrow n_1 - 1, \gamma 0, Z}^{-1}
\end{align}
For the total time required for pumping from an $n_1$ to an $n_1^{'}$ we add the average times for the intermediate steps
\begin{align}
\tau_{n_1 \rightarrow n_1^{'}}
= \sum_{n = n_1^{'} + 1}^{n_1} \tau_{n \rightarrow n - 1, \gamma 0, Z}
= \sum_{n = n_1^{'} + 1}^{n_1} \Gamma_{n \rightarrow n - 1, \gamma 0, Z}^{-1}
= \frac{(\gamma_e + \kappa_b)^2}{2 \gamma_{0e}} \sum_{n = n_1^{'} + 1}^{n_1} \frac{1}{n (\Omega_Z^{(F=n)})^2}
\label{taun1ton1minus1}
\end{align}
We can also assign a total decay rate from an initial state with $n_1$ to a final state with $n_1^{'}$, $\Gamma_{n_1 \rightarrow n_1^{'}} = 1/\tau$. It is however more useful to use the total preparation time and to minimize it by the choice of available parameters. Here, in particular the Rabi frequencies $\Omega_{\rm Z}^{(F)}$ of individual field tones $F$ can be chosen, as well as the tunable decay rate $\gamma_e$.

Since the pumping occurs from a maximal $n_1$ of $N-1$ to $n_1 - 1$, the worst case preparation time for Z pumping is (using Eq. \eqref{taun1ton1minus1}) found to be given by
\begin{align}
\tau_{n_1=N-1 \rightarrow n_1=0}
= \sum_{n = 1}^{N-1} \tau_{n \rightarrow n - 1, \gamma 0, Z}
= \sum_{n = 1}^{N-1} \Gamma_{n \rightarrow n - 1, \gamma 0, Z}^{-1}
= \frac{(\gamma_e + \kappa_b)^2}{2 \gamma_{0e}} \sum_{n_1 = 1}^{N-1} \frac{1}{n_1  (\Omega_Z^{(F=n_1)})^2}
\label{taufromj0}
\end{align}
The time from $n_1 = 1$ to $\ket{\rm GHZ}$ differs from that to $n_1 = 0$ by a factor of $2$. This is due to the fact that only half of the population is pumped to $\ket{\rm GHZ}$ and the other half to $\ket{\rm GHZ_-}$, which is continuously depumped by the X pumping discussed below. Therefore, on average two attempts are required so that the preparation time is doubled,
\begin{align}
\tau_{n_1=N-1 \rightarrow {\rm GHZ}} = 2 \tau_{n_1=N-1 \rightarrow n_1=0}
= \frac{(\gamma_e + \kappa_b)^2}{\gamma_{0e}} \sum_{n_1 = 1}^{N-1} \frac{1}{n_1  (\Omega_Z^{(F=n_1)})^2}
\label{worstcaseZpreptimewithfactor2}
\end{align}

As anticipated already before Eq.\ (\ref{gammaminusformula}), for GHZ preparation we will from now on choose the parameter values $\kappa_b=\kappa_c=0$, $\gamma_{0e}=\gamma_{1e}=\gamma_e/2$, and abbreviate $\gamma_e\equiv\gamma$. In order to obtain dimensionless optimization variables, we will furthermore write $\Omega_Z^{(F=n_1)}=:A_{F}\Omega$ (for $F=1,2,\ldots,N-1$) with nonnegative dimensionless variables $A_{F}$. The quantity $\Omega$ is a dimensionful frequency parameter, whose size has been chosen such that (for the weak driving calculation) all $\Omega_Z^{(F=n_1)}$ have to satisfy $\Omega_Z^{(F=n_1)}\leq\Omega$, i.e.\ such that the dimensionless parameters $A_{F}$ satisfy $A_{F}\in[0,1]$. In this notation, the above GHZ preparation time reads:
\begin{align}
\tau_{n_1=N-1 \rightarrow {\rm GHZ}} = \frac{2\gamma}{\Omega^2} \sum_{F = 1}^{N-1} \frac{1}{F A^2_{F}}=\frac{2\gamma}{\Omega^2}H(\{A_{F}\})~,\label{restatetaupessimist}
\end{align}
where we have defined the function
\begin{align}
H(\{A_{F}\})~&:=~\sum_{F=1}^{N-1} \frac{1}{F A_{F}^2}~.\label{firsttimeL}
\end{align}

With the same abbreviations, the error rate from Z pumping alone reads, from Eq.\ (\ref{gammaminusformula}):
\begin{align}
\Gamma_{{\rm GHZ} \rightarrow ?, \gamma, Z} &= \frac{3 \gamma_e N}{16 g^2} \sum_{F=1}^{N-1} F \left(\frac{\Omega_Z^{(F)}}{N - F}\right)^2
=\frac{3 \gamma\Omega^2}{16 g^2} N \sum_{F=1}^{N-1}\frac{F A_{F}^2}{(N - F)^2} = \frac{3 \gamma\Omega^2}{16 g^2} N G(\{A_F\})\label{restateGammaminus},
\end{align}
where
\begin{align}\label{asdfasdfzuio}
G(\{A_{F}\})~&:=~\sum_{F=1}^{N-1}\frac{F A_{F}^2}{(N-F)^2}~.
\end{align}

Our goal for now is to find parameters $\{A_{F}\}$ that minimize the $Z$-error for any given value of the GHZ preparation time (or, equivalently, minimize the GHZ preparation time for any fixed error value). Minimizing $G(\{A_{F}\})$ under the constraint $H(\{A_{F}\})\equiv\widetilde{H}\equiv{\rm const}$ by the method of Lagrange multipliers leads to
\begin{align}
A_{F}^2~=~\eta\frac{N-F}{F}\,,\quad\text{with}~\,\eta>0~\,\text{such that}~~~\frac{1}{\eta}\sum_{F=1}^{N-1}\frac{1}{N-F}~=~\widetilde{H}~.
\end{align}
Approximating the latter harmonic sum gives roughly $\eta\approx(\log N)/\widetilde{H}$, and we would have $A_{F}>1$ for some $F$ (in particular, for $F=1$) if $\eta>1/(N-1)$, i.e.\ if $\widetilde{H}\lesssim(N-1)\log N$.

However, one can still find the optimal assignment $\{A_{F}\}$ minimizing $G(\{A_{F}\})$ while obeying $0\leq A_{F}\leq1$ and $H(\{A_{F}\})\leq\widetilde{H}$:
\begin{align}\label{Afceilingassingment}
A_{F}^2~=~\left\lceil\eta\frac{N-F}{F}\right\rceil^1~,
\end{align}
where we defined the ``ceil-1'' function
\begin{align}\label{ceil1function}
\lceil x\rceil^1~:=~\left\{\begin{matrix}1\quad&\text{if}~x>1\\x\quad&\text{if}~x\leq1\end{matrix}\right.~,
\end{align}
and $\eta$ needs to be adjusted such that $H(\{A_{F}\})=\widetilde{H}$. The assignment (\ref{Afceilingassingment}) means that $A_{F}^2=1$ for $F<N/(1+1/\eta)$ and $A_{F}^2=\eta(N-F)/F$ for $F\geq N/(1+1/\eta)$. That (\ref{Afceilingassingment}) is the unique optimal solution can be checked by the Karush-Kuhn-Tucker (KKT) conditions \cite[Section 5.3.3]{boydconvex}, using that the function $G(\{A_{F}\})$ to be minimized and the constraint function $H(\{A_{F}\})$ are both strictly convex in their arguments.

For $\eta\in[1/(N-1),N-1]$ (such that the selection between the two cases in (\ref{ceil1function}) happens at some $F\in[1,N-1]$) one can thus compute:
\begin{align}
H(\{A_F\})~&=~\sum_{F=1}^{N-1} \frac{1}{A_F^2\, F}~\approx~\sum_{F=1}^{N/(1+1/\eta)}\frac{1}{F}\,+\,\sum_{F=N/(1+1/\eta)}^{N-1}\frac{1}{\eta(N-F)}\\
&\approx~\log\frac{N}{1+1/\eta}\,+\,\frac{1}{\eta}\log N\left(1-\frac{1}{1+1/\eta}\right)\\
&\approx~\left(1+\frac{1}{\eta}\right)\log N~+~\frac{1}{\eta}\log\frac{1}{\eta}-\left(1+\frac{1}{\eta}\right)\log\left(1+\frac{1}{\eta}\right)\\
&\approx~\left(1+\frac{1}{\eta}\right)\log N~,\label{Hforgeneralconsteta}
\end{align}
and
\begin{align}
G(\{A_F\})~&=~\sum_{F=1}^{N-1}\frac{A_F^2\,F}{(N-F)^2}~\approx~\sum_{F=N/(1+1/\eta)}^{N-1}\frac{\eta}{N-F}\,+\,\sum_{F=1}^{N/(1+1/\eta)}\frac{F}{(N-F)^2}\\
&\approx~\eta\log N\left(1-\frac{1}{1+1/\eta}\right)\,+\,\eta\,+\,\log\left(1-\frac{1}{1+1/\eta}\right)\\
&\approx~\eta\log N\,+\,\eta\,-\,(\eta+1)\log(\eta+1)\\
&\approx~\eta\log N~.\label{Gforgeneralconsteta}
\end{align}
For $\eta={\rm const}>0$, the error in both estimates is $\mathcal{O}(1)$ as $N\to\infty$ and thus irrelevant compared to the $\log N$ terms. In particular, if one wants the numerical factor $H$ in the preparation time (\ref{restatetaupessimist}) to scale like $\log N$ (like optical pumping), then one needs $\eta\geq \mathcal{O}(1)$, which we achieve by letting $\eta={\rm const}$ as $N\to\infty$.

The error $E$ can generally be obtained by comparing the preparation time of the desired state of the protocol and the loss rate out of it. Using Eqs. (\ref{restatetaupessimist}) and (\ref{restateGammaminus}), the error $E$ is found to be given by
\begin{align}\label{errorinZpumping}
E~:=~(\tau_{n_1=N-1 \rightarrow {\rm GHZ}})(\Gamma_{{\rm GHZ} \rightarrow ?, \gamma, Z})~=~\frac{3\gamma^2}{8g^2}\,N\,G(\{A_F\})\,H(\{A_F\})~.
\end{align}
Thus, to achieve a desired error $E$ (which may be $N$-dependent, i.e.\ $E=E(N)$), one can adjust $\gamma$ appropriately. If we assume a relation $\Omega=\alpha\gamma$ in order to limit $\Omega$ to the weak driving regime (and where the number $\alpha$ may or may not depend on $N$, i.e.\ $\alpha=\alpha(N)$), we can plug this back into the worst-case preparation time (\ref{restatetaupessimist}) to obtain
\begin{align}
\tau_{n_1=N-1 \rightarrow \rm GHZ}~=~\frac{2\gamma}{\Omega^2}H~=~\frac{2}{\alpha^2\gamma}H~&=~\sqrt{\frac{3}{2}}\frac{1}{\sqrt{E}\alpha^2g}\sqrt{N\,G\,H^3}\\
&\approx~\eta^{1/2}\left(1+\frac{1}{\eta}\right)^{3/2}\frac{\sqrt{3/2}}{\sqrt{E}\alpha^2g}\,\,N^{1/2}\log^2N~,\label{beforedeterminingoptimaleta}
\end{align}
where the last estimate holds for $\eta={\rm const}$ as $N\to\infty$ and we have neglected lower-order terms in $N$. The prefactor is minimized for $\eta=2$, leading to a minimal preparation time (for the desired error $E$):
\begin{align}
\tau_{n_1=N-1 \rightarrow \rm GHZ}~\approx~\frac{9}{2\sqrt{2}}\frac{1}{\sqrt{E}\alpha^2g}\,\,N^{1/2}\log^2N~\approx~\frac{3.2}{\sqrt{E}\alpha^2g}\,\,N^{1/2}\log^2N~.\label{tau1Nforoptimalconsteta2}
\end{align}

\bigskip

To summarize the optimal parameter choices for the scenario considered here:

\begin{itemize}
\item The $A_F$ for $1\leq F\leq N-1$ have to be chosen as follows (cf.\ Eq.\ (\ref{Afceilingassingment}) with the optimal choice $\eta=2$):
\begin{align}A_F=\sqrt{\lceil2(N-F)/F\rceil^1},\label{finalresultforAfZpumping}\end{align}i.e.\ $A_F=1$ for $1\leq F\leq2N/3$, and $A_F=\sqrt{2(N-F)/F}$ for $2N/3<F\leq N-1$. This leads to:
\begin{align}
H(\{A_F\})~&=~\frac{3}{2}\log N\,+\,\mathcal{O}(1)~,\label{timescalingfixedEforeta2test}\\
G(\{A_F\})~&=~2\log N\,+\,\mathcal{O}(1)~.\label{timescalingfixedEforeta2testtext}
\end{align}
\item The choice of $\gamma\equiv\gamma_e$, and consequently of $\Omega\equiv\alpha\gamma$ and $\gamma_{0e}=\gamma_{1e}=\gamma/2$, is given by:
\begin{align}
\gamma=\frac{2\sqrt{2}}{3}\frac{g\sqrt{E}}{\sqrt{N}\log N}\,.\label{gammafinalfromGHZ-Zpumpingonly}
\end{align}
We also set $\kappa_b=\kappa_c=0$. This leads to:
\begin{align}
\tau_{n_1=N-1 \rightarrow \rm GHZ}~&=~\frac{9}{2\sqrt{2}}\frac{1}{\sqrt{E}\alpha^2g}\,N^{1/2}\log^2N~.\label{timescalingfixedEforeta2}
\end{align}
The appearance of $N^{1/2}$ in the scaling of $\tau_{n_1=N-1 \rightarrow \rm GHZ}$ can be traced back to the fact that the noise $\Gamma_-$ acts on each of the $N$ atoms (i.e.\ the prefactor of $N$ in Eq.\ (\ref{gammaminusformula})).
\end{itemize}

\begin{figure}[t]
\centering
\includegraphics[scale=1.3]{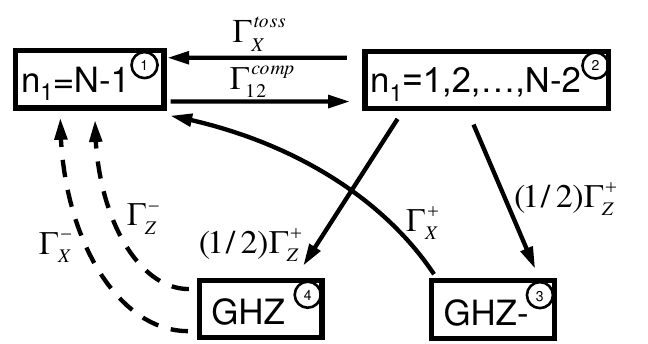}
\caption{\label{comp4pessfig}Four-compartment-model of the process creating $\ket{\rm GHZ}$.}
\end{figure}

\subsection{Compartment model and effective rates}
\label{subsectioncompartmentmodel}

While so far we have only discussed the Z pumping, we will now make a simplified model of both processes, Z and X, that create the $\ket{\rm GHZ}$ state. This model is more pessimistic w.r.t.\ the preparation time and the error treatment than the actual (Lindbladian) dynamics described in Section \ref{SecEngineeredGHZ}. Nevertheless, this model will still yield a scaling of the preparation time as $\tau_{\rm GHZ}\sim\!(N^{1/2}\log^2N)/\sqrt{E}$ with the number of qubits $N$ and the error $E$, just like Eq.\ (\ref{tau1Nforoptimalconsteta2}), which thus shows that this is indeed the best achievable scaling.

Our simplified model is shown in Fig.\ \ref{comp4pessfig} and consists of four compartments as we explain now. We split the $2^N$-dimensional Hilbert space into the subspaces with $n_1=1,2,\ldots,N-1$, counting the number of $\ket{1}$-states appearing in a computational basis state (as in Section \ref{SecEngineeredGHZ}). All states with $n_1=N-1$ are put into compartment 1, as they are furthest away (in pumping time) from the GHZ states; the states with $n_1=1,\ldots,N-2$ are then put into compartment 2. The two Hilbert space dimensions belonging to $n_1=0$ and to $n_1=N$ are split into the compartments $3$ and $4$ of Fig.\ \ref{comp4pessfig}, corresponding to the $\ket{\rm GHZ_-}$ and the $\ket{\rm GHZ}$ states, respectively. The desired Z pumping process creates each of those GHZ states with equal rate $\frac{1}{2}\Gamma_Z^+$ out of compartment 2 where using Eq.\ (\ref{taun1ton1minus1}) we find
\begin{align}
\Gamma_Z^+&=(\tau_{n_1=N-2 \rightarrow n_1=0})^{-1}=\left(\frac{\gamma}{\Omega^2}\sum_{F=1}^{N-2}\frac{1}{F A_F^2}\right)^{-1}=\left(\frac{3\gamma}{2\Omega^2}\log N\right)^{-1}=\frac{\Omega^2}{\gamma}\frac{2}{3\log N}\,.
\end{align}
Here, we have again plugged in $\gamma_{0e}=\gamma/2$, $\kappa_b=0$ and the optimal $A_j$ from Eq.\ (\ref{finalresultforAfZpumping}) and followed the same computation as for (\ref{Hforgeneralconsteta}), neglecting lower-order terms. Similarly, the rate from compartment 1 to compartment 2 is given by Eq.\ (\ref{rateasdfasdfasdfabasdbasdbasb}):
\begin{align}
\Gamma^{comp}_{12}=\Gamma_{n_1=N-1\to n_1=N-2,\gamma0,Z}^{(F=N-1)}\approx\frac{2 (N-1) \gamma_{0e} (\Omega_Z^{(F=N-1)})^2}{(\gamma_e + \kappa_b)^2} = \frac{2\Omega^2}{\gamma}.
\end{align}

As compartment 1 is the furthest away from the desired GHZ state, we pessimistically model the $X$-toss (i.e.\ the action of the X process on the states other than $\ket{\rm GHZ}$ and $\ket{\rm GHZ_-}$, see Eq. \eqref{Xtossrate}) to throw any state back to the $n_1=N-1$ compartment, with a rate $\Gamma_X^{toss}$ to be computed below. Similarly, we model the errors $\Gamma_Z^-$  and $\Gamma_X^-$ affecting $\ket{\rm GHZ}$ such that they move the $\ket{\rm GHZ}$-state back to compartment $1$. By the same rationale, even the ``good'' X process $\Gamma_X^+$ is modelled to take the un-wanted state $\ket{\rm GHZ_-}$ back to compartment 1.
\\
The detrimental Z rate for the compartment model as computed with the parameters from the previous subsection, Eq.\ (\ref{finalresultforAfZpumping}) (see also Eqs.\ (\ref{restateGammaminus}), (\ref{asdfasdfzuio}), and (\ref{Gforgeneralconsteta}) with the optimal $\eta=2$) is
\begin{align}
\Gamma^-_Z=\Gamma_{{\rm GHZ} \rightarrow ?, \gamma, Z} = \frac{3 \gamma\Omega^2}{16 g^2} N G(\{A_f\})=\frac{\gamma\Omega^2}{g^2}\frac{3N\log N}{8}.
\end{align}

For the X rates, it was found in Section \ref{subsectiondepumpingGHZX} that only the Rabi oscillations $\Omega_X^{(F)}$ with \emph{odd} index $F$ should be turned on. Similar to Section \ref{SecPreparation}, we write $\Omega_X^{(F)}=A_X^{(F)}\Omega$ with the dimensionless parameters $A_X^{(F)}$ which we discuss below. The ``good'' X rate $\Gamma^+_X$ for the compartment model of Fig.\ \ref{comp4pessfig} is then given by Eq.\ (\ref{GammaXtotal}):
\begin{align}\label{startcomputinggoodXrateasdfqwerasdf}
\Gamma^+_X&=\Gamma_{{\rm GHZ_-} \rightarrow ?, X}
\approx \frac{2 \gamma_f}{(\gamma_f + \kappa_c)^2} \sum_{F=1,3,...}^N {N \choose F} \frac{F (\Omega_{X}^{(F)})^2}{2^{N-1}}\\
&=\frac{2\Omega^2}{\gamma}\sum_{F=1,3,\ldots}^N\frac{1}{2^{N-1}}{N\choose F}F(A_X^{(F)})^2.
\end{align}
Since the (normalized) probability distribution $\{{N\choose F}/2^{N-1}\}_{F=1,3,\ldots}^N$ is strongly peaked around the values $F\sim N/2$ (similar to the full binomial distribution), the exact functional choice of the X coefficients $A_X^{(F)}$ (for odd $F$) does not really matter in the limit of large $N$, as long as it is not exponentially fine tuned. Since a similar binomial distribution occurs in the X error rates below as well and as the same reasoning applies, we can take all $A_X^{(F)}$ to be equal as a very good approximation:
\begin{align}
A_X^{(F=1)}=A_X^{(F=3)}=A_X^{(F=5)}=\ldots\equiv A_X.
\end{align}
With this, the rate $\Gamma^+_X$ evaluates to (with exponentially good accuracy for large $N$):
\begin{align}
\Gamma^+_X=\frac{2\Omega^2}{\gamma}\,\frac{N}{2}\,A_X^2=\frac{\Omega^2}{\gamma}NA_X^2,\label{goodXrate2}
\end{align}
As we will see below, with this choice the error induced by the X error rate is smaller than the one from the Z process and thus not a major limitation to the preparation of the entangled state.

The X error rate $\Gamma_X^-$ is given by (\ref{GammaXminus}), and we simplify it again with our above parameter choices, making approximations of the binomial distribution and the other sum which are good in the large-$N$ limit:
\begin{align}
\Gamma^-_X&=\Gamma_{{\rm GHZ} \rightarrow ?, \gamma, X} \approx \frac{\gamma_{f}}{2 g^2} \sum_{n_=0,2,...}\frac{1}{2^{N-1}} {N \choose n_-} n_- \sum_{F=1,3,...} F \left(\frac{\Omega_X^{(F)}}{F - n_-}\right)^2\\
&\approx\frac{\gamma\Omega^2 A_X^2}{2g^2}\,\frac{N}{2}\,\sum_{F=1,3,\ldots}\frac{F}{\left(F-[N/2]_{even}\right)^2}.
\end{align}
In the expression above, $[N/2]_{even}$ denotes the next higher integer number of $N/2$. Having the limit of large $N$ in mind, we evaluate the last sum as follows:
\begin{align}
\sum_{F=1,3,\ldots}&\frac{F}{\left(F-[N/2]_{even}\right)^2}=\sum_{1\leq F\leq[N/2]_{even}-1}^{F odd}\frac{F}{\left(F-[N/2]_{even}\right)^2}+\sum_{[N/2]_{even}+1\leq F\leq N}^{F odd}\frac{F}{\left(F-[N/2]_{even}\right)^2}\\
=&\sum_{1\leq F\leq[N/2]_{even}-1}^{F odd}\frac{[N/2]_{even}}{\left(F-[N/2]_{even}\right)^2}+\sum_{1\leq F\leq[N/2]_{even}-1}^{F odd}\frac{F-[N/2]_{even}}{\left(F-[N/2]_{even}\right)^2}\\
&+\sum_{[N/2]_{even}+1\leq F\leq N}^{F odd}\frac{[N/2]_{even}}{\left(F-[N/2]_{even}\right)^2}+\sum_{[N/2]_{even}+1\leq F\leq N}^{F odd}\frac{F-[N/2]_{even}}{\left(F-[N/2]_{even}\right)^2}\\
=&[N/2]_{even}\left(\frac{1}{1^2}+\frac{1}{3^2}+\ldots+\frac{1}{([N/2]_{even}-1)^2}\right)-\left(\frac{1}{1}+\frac{1}{3}+\ldots+\frac{1}{([N/2]_{even}-1)}\right)\\
&+[N/2]_{even}\left(\frac{1}{1^2}+\frac{1}{3^2}+\ldots+\frac{1}{(N-[N/2]_{even}-1)^2}\right)+\left(\frac{1}{1}+\frac{1}{3}+\ldots+\frac{1}{(N-[N/2]_{even}-1)}\right)\\
\approx&2\cdot\frac{N}{2}\sum_{n=1,3,5,\ldots}^{\infty}\frac{1}{n^2}=2\cdot\frac{N}{2}\cdot\frac{3}{4}\sum_{n=1,2,3,4,\ldots}^\infty\frac{1}{n^2}=N\frac{3\pi^2}{4\cdot6}\approx1.23N\approx\frac{5N}{4},
\end{align}
neglecting subleading terms in $N$. This finally gives:
\begin{align}
\Gamma_X^-=\frac{\gamma\Omega^2}{g^2}\,\frac{5N^2 A_X^2}{16}.\label{badXrate}
\end{align}

Finally, the X toss rate $\Gamma_X^{toss}$ in Fig.\ \ref{comp4pessfig} is given by Eq.\ (\ref{Xtossrate}), which with our parameter choices becomes [note that (\ref{Xtossrate}) is half of the ``good'' rate (\ref{GammaXtotal}), which we have computed in Eq.\ (\ref{goodXrate2}) already]:
\begin{align}
\Gamma^{toss}_X=\frac{\Omega^2}{\gamma}\frac{N A_X^2}{2}.\label{Xtossrate2}
\end{align}
From the process in Fig.\ \ref{comp4pessfig} one can see that the optimal parameters $A_X$ have to be chosen such that the good $X$-process and the good $Z$-process have about the same rate:
If the X pumping rate $\Gamma^+_X$ is too weak, population will accumulate in $\ket{\rm GHZ_-}$ by the Z pumping ($\Gamma_Z^+$). On the other hand, a too strong X pumping will hinder the preparation mechanism through the X toss effect. We thus set the rates for the desired processes, Z pumping and X depumping to be equal,
\begin{align}
\Gamma_X^+ = \Gamma_Z^+.
\end{align}
This results in:
\begin{align}
A_X^2=\frac{2}{3N\log N}\label{finalchoiceAX}
\end{align}
(note that, for all $N\geq2$, this choice is consistent with the requirement $A_X\leq1$ for the weak-driving analysis).

With these choices made, we summarize the parameters and effective rates for the four-compartment model of Fig.\ \ref{comp4pessfig} found so far:
\begin{itemize}
\item For the $Z$-pumping, we make the choice of parameters found to be optimal in Eq.\ (\ref{finalresultforAfZpumping}):
\begin{align}
A^{(F)}_Z~=~\left\{\begin{array}{ll}1&\text{for}~F\leq2N/3~~,\\\sqrt{2\frac{N-F}{F}}\quad&\text{for}~F\geq2N/3~,\end{array}\right.\label{laliluasdfqwer}
\end{align}
meaning that $\Omega_Z^{(F)}=\Omega A^{(F)}_Z$.
\item For the $X$-pumping we take (see Eq.\ (\ref{finalchoiceAX})):
\begin{align}
A_X^{(F=odd)}~=~\sqrt{\frac{2}{3}\,\frac{1}{N\log N}}~,\qquad A_X^{(F=even)}=0~.\label{laliluasdfqwerasdfqwer}
\end{align}
\item Then one obtains for the effective rates in Fig.\ \ref{comp4pessfig}:
\begin{align}\label{listgoodgammasinpess3compmodel}
\Gamma_Z^+~=~\Gamma_X^+~&=~\frac{\Omega^2}{\gamma}\,\frac{2}{3\log N}~,\\
\Gamma_{12}^{comp}~&=~\frac{\Omega^2}{\gamma}\cdot2~=~\Gamma_Z^+\cdot3\log N,\\
\Gamma_{X}^{toss}~&=~\frac{\Omega^2}{\gamma}\,\frac{1}{3\log N}~=~\frac{1}{2}\Gamma_Z^+~,\\
\Gamma_Z^-~&=~\frac{\gamma\Omega^2}{g^2}\frac{3N\log N}{8}~,\\
\Gamma_X^-~&=~\frac{\gamma\Omega^2}{g^2}\,\frac{5N}{24\log N}~.
\end{align}
\item The total error rate (leading from compartment 4 to compartment 1) is thus
\begin{align}
\Gamma_-~:=~\Gamma_Z^-+\Gamma_X^-~&=~\frac{\gamma\Omega^2}{g^2}\,\frac{3N\log N}{8}\,\left(1+ \frac{5}{9\log^2N}\right)~=~\Gamma_Z^+\cdot\frac{\gamma^2}{g^2}\frac{9N\log^2N}{16}\,\left(1+ \frac{5}{9\log^2N}\right)\label{extreactgammadivg}.
\end{align}
\end{itemize}

\subsection{Transition matrix, stationary error, and GHZ preparation time}\label{subsectiontransitionmatrixetc}
The transition matrix for the 4-compartment model of Fig.\ \ref{comp4pessfig} is:
\begin{align}\label{transitionmatrix4x4pess}
T~=~\begin{pmatrix}-\Gamma_{12}^{comp}&\Gamma_X^{toss}&\Gamma_X^+&\Gamma_-\\\Gamma_{12}^{comp}&-(\Gamma_X^{toss}+\Gamma_Z^+)&0&0\\0&\frac{1}{2}\Gamma_Z^+&-\Gamma_X^+&0\\0&\frac{1}{2}\Gamma_Z^+&0&-\Gamma_-\end{pmatrix}~
=~\Gamma_Z^+\begin{pmatrix}-3\log N&\frac{1}{2}&1&\Gamma_-/\Gamma_Z^+\\3\log N&-\frac{3}{2}&0&0\\0&\frac{1}{2}&-1&0\\0&\frac{1}{2}&0&-\Gamma_-/\Gamma_Z^+\end{pmatrix}~.
\end{align}
The steady-state population $p_\infty:=(P_1(\infty),P_2(\infty),P_3(\infty),P_4(\infty))$ is given as the solution (normalized to the sum of entries being 1) of the equation $Tp_\infty=0$. This gives:
\begin{align}
p_\infty~=~\left(\begin{array}{l}P_1(\infty)\\P_2(\infty)\\P_3(\infty)\\P_4(\infty)\end{array}\right)~=~\left(\begin{array}{c}1/\log N\\2\\1\\\Gamma_Z^+/\Gamma_-\end{array}\right)\cdot\frac{1}{3+\frac{1}{\log N}+\frac{\Gamma_Z^+}{\Gamma_-}}~.
\end{align}
The \emph{steady-state fidelity} is just $F=P_4(\infty)$, and the \emph{error} is thus
\begin{align}
E~&=~1-F~=~1-P_4(\infty)~=~1-\frac{1}{1+\frac{\Gamma_-}{\Gamma_Z^+}\left(3+\frac{1}{\log N}\right)}\label{exactfidelitypess4compmodel}\\
&\approx~\frac{\Gamma_-}{\Gamma_Z^+}\left(3+\frac{1}{\log N}\right)~=~\frac{\gamma^2}{g^2}\frac{27N\log^2N}{16}\,\left(1+ \frac{5}{9\log^2N}\right)\left(1+\frac{1}{3\log N}\right)~.\label{approximatedfidelitypess4compmodel}
\end{align}
(Here the approximation in the second line was made for analytical convenience and gives a slightly pessimistic bound.) In the scaling with large $N$, this expression for $E$ agrees with the one implied by Eq.\ (\ref{gammafinalfromGHZ-Zpumpingonly}) that was found by other means before, and the prefactor is similar. Thus, to achieve a desired stationary error $E$, we need to adjust $\gamma$ such that:
\begin{align}
\gamma~=~g\sqrt{E}\left[\frac{27N\log^2N}{16}\,\left(1+ \frac{5}{9\log^2N}\right)\left(1+\frac{1}{3\log N}\right)\right]^{-1/2}\,.\label{gammawithXandZpumpingasdfqer}
\end{align}
Below we will use this expression instead of the results obtained in (\ref{gammafinalfromGHZ-Zpumpingonly}) which were derived by considered only the Z pumping.

So far we have discussed the Z pumping separately from the X pumping, deriving an individual characteristic time $\tau_{n_1=N-1 \rightarrow {\rm GHZ}}$. It now remains to derive an analytical expression for the total GHZ pumping time that is obtained in the presence of X pumping, using the parameters (\ref{gammawithXandZpumpingasdfqer}) and (\ref{finalresultforAfZpumping}). In order to factor out the dependence of the stationary error $E$ and to obtain a tractable analytical expression, we make (for the computation of the GHZ preparation time) an approximation to the transition matrix (\ref{transitionmatrix4x4pess}) by dropping the small terms leading out of the GHZ state (these terms vanish in the limit of $E\to0$). That is, we set the fourth column in (\ref{transitionmatrix4x4pess}) to zero:
\begin{align}\label{transitionmatrix4x4pessgood}
T_+~=~\Gamma_Z^+\begin{pmatrix}-3\log N&\frac{1}{2}&1&0\\3\log N&-\frac{3}{2}&0&0\\0&\frac{1}{2}&-1&0\\0&\frac{1}{2}&0&0\end{pmatrix}~.
\end{align}
For the initial population vector $(P_1(0),P_2(0),P_3(0),P_4(0))=(0,0,1,0)$, which corresponds to the whole population being in the worst state $\ket{\rm GHZ_-}$ of Fig.\ \ref{comp4pessfig}, we have the following evolution:
\begin{align}
\begin{pmatrix}P_1(t)\\P_2(t)\\P_3(t)\\P_4(t)\end{pmatrix}~=~e^{tT_+}\begin{pmatrix}P_1(0)\\P_2(0)\\P_3(0)\\P_4(0)\end{pmatrix}~=~e^{(t\Gamma_Z^+)(T_+/\Gamma_Z^+)}\begin{pmatrix}0\\0\\1\\0\end{pmatrix}~.\label{evolutionwithapproxTplus}
\end{align}
Thus, on time-scales larger than any fixed $t_0$, the transition from the worst state $\ket{\rm GHZ_-}$ to the desired state $\ket{\rm GHZ}$ happens at least as fast as in an exponential decay process, with an approximation $\Gamma_+$ for the effective exponential rate computed as:
\begin{align}\label{rateasdfqwerycxvasdfqutqw}
P_4(t_0)~=~1-e^{-\Gamma_+ t_0}\,,\quad\text{i.e.}\quad\Gamma_+~=~\Gamma_Z^+\,\frac{-\log\left(1-P_4(t_0)\right)}{\Gamma_Z^+ t_0}~.
\end{align}
Note that the fraction in the last expression will depend both on $(\Gamma_Z^+ t_0)$ and on $N$, since the $N$-dependence of the transition matrix $T_+$ in Eq.\ (\ref{transitionmatrix4x4pessgood}) cannot be factored out completely. To get a meaningful expression for the exponential rate, the timescale $t_0$ should be chosen comparable to the other relevant timescales of the process. For definiteness we will thus set $t_0:=1/\Gamma_Z^+$ throughout.

The characteristic preparation time of $\ket{\rm GHZ}$ in the sense of an exponential rate is then:
\begin{align}\label{firsteqnasdfqweryxcv}
\tau_{\rm GHZ}~&=~\frac{1}{\Gamma_+}~=~\frac{1}{\Gamma_Z^+}\left(\frac{-\log\left(1-P_4(t_0)\right)}{\Gamma_Z^+t_0}\right)^{-1}~\stackrel{[\text{Eq.}\ (\ref{listgoodgammasinpess3compmodel})]}{=}~\frac{3\gamma}{2\Omega^2}\log N\,\left(\frac{-\log\left(1-P_4(t_0)\right)}{\Gamma_Z^+t_0}\right)^{-1}\\
&\stackrel{[\Omega=\alpha\gamma]}{=}~\frac{3}{2\alpha^2\gamma}\log N\,\left(\frac{-\log\left(1-P_4(t_0)\right)}{\Gamma_Z^+t_0}\right)^{-1}\\
&\stackrel{[\text{Eq.}~(\ref{gammawithXandZpumpingasdfqer})]}{=}~\frac{N^{1/2}\log^2N}{\alpha^2g\sqrt{E}}\,\left[\frac{9\sqrt{3}}{8}\,\frac{\Gamma_Z^+t_0}{-\log\left(1-P_4(t_0)\right)}\,\sqrt{\left(1+ \frac{5}{9\log^2N}\right)\left(1+\frac{1}{3\log N}\right)}\,\right]\,.
\label{upperboundontauGHZinpess4compmodel}
\end{align}
In the limit of large $N$, the square root inside the bracketed expression will tend to $1$. And also the fractional expression involving $P_4(t_0)$ will tend to a constant number independent of $N$, since large values of the first column in the matrix in (\ref{transitionmatrix4x4pessgood}) mean that the transition time out of the first compartment is insignificant compared to the other  transition times, which are all \emph{in}dependent of $N$. This is also easily seen numerically.

In the following table, we evaluate the factor in square brackets (which we call $b(N)$) in Eq.\ (\ref{upperboundontauGHZinpess4compmodel}) for different values of $N$:
\begin{center}
\begin{tabular}{|l||l|l|l|l|l|l|l|l|l|l|l|l|l|l|l|l|l|}
\hline
$N$ & $2$ & $3$ & $4$ & $5$  & $6$ & $7$ & $8$ & $10$ & $20$ & $50$ & $100$ & $500$ & $1000$  & $10^4$ & $10^5$ & $10^6$ \\
\hline
$b(N)$ & 55 & 33 &27 & 25 &  23 & 22 & 21 & 20 & 18 & 17 & 16 & 15.5 & 15 & 14.6 & 14.3 & 14.1 \\
\hline
\end{tabular}
\end{center}

The GHZ preparation time is then:
\begin{align}
\tau_{\rm GHZ}~=~b(N)\frac{\sqrt{N}\log^2N}{\alpha^2g\sqrt{E}}\,,\label{GHZwithXandZpreptimewithprefbN}
\end{align}
where $b(N)$ tends to about $13$ as $N\to\infty$ (see table above). The parameter choices for this can be found in Eqs.\ (\ref{laliluasdfqwer}), (\ref{laliluasdfqwerasdfqwer}), and (\ref{gammawithXandZpumpingasdfqer}) together with $\Omega=\alpha\gamma$, $\gamma_{0e}=\gamma_{1e}=\gamma$, $\kappa_b=\kappa_c=0$.

\subsection{Analysis of the ``dynamical problem''}
\label{AppDynamical}

Here we analyze the ``dynamical problem''. This means that, for a certain exponential form of the time-evolution of the GHZ error $E(t)$ (or, equivalently, of the GHZ fidelity $F(t)=1-E(t)$) in an effective model, we compute and minimize the time $t$ it takes to achieve a desired target error ${\mathcal E}$. The main result will be that, up to an additional factor of $\log(1/{\mathcal E})$, the scaling of the preparation time with the particle number $N$ and with the error ${\mathcal E}$ is the same as in the previous analytical approaches, see e.g.\ Eqs.\ (\ref{timescalingfixedEforeta2}) and (\ref{GHZwithXandZpreptimewithprefbN}) (with $E$ replaced by ${\mathcal E}$).

In the following, we write again $\Omega=\alpha\gamma$ to be able to limit our treatment to the weak-driving regime by suitable choices of the number $\alpha$. Furthermore, to keep the main derivation as general as possible, we write the rates into and out of the desired GHZ state as
\begin{align}
\Gamma~&=~\frac{\Omega^2}{\gamma}f(N)~=~\alpha^2\gamma f(N)~,\label{Gammaindynamicalsectionasdf}\\
\Gamma_-~&=~\frac{\gamma\Omega^2}{g^2}h(N)~\label{Gammaminuswithh}
\end{align}
with functions $f\equiv f(N)$ and $h\equiv h(N)$. Later, we will evaluate our results for the functions
\begin{align}
h(N)~=~\frac{3N\log N}{8}\,\left(1+ \frac{5}{9\log^2N}\right)\,,\label{choiceforhNasdfqwer}
\end{align}
which is motivated by Eq.\ (\ref{extreactgammadivg}), and
\begin{align}
f(N)~=~\frac{2}{3+9\log N}\,\,,\label{choiceforfNasdfqwer}
\end{align}
which is chosen such that the ratio $\Gamma_-/\Gamma$ yields the stationary error from Eq.\ (\ref{approximatedfidelitypess4compmodel}).

Finally, we make the following basic ansatz for the time evolution of the error:
\begin{align}
E(t)~&=~\frac{\Gamma_-}{\Gamma}\,+\,\left(1-\frac{\Gamma_-}{\Gamma}\right)e^{-t\kappa\Gamma}\,.\label{effectivedecayindynamicalproblemasdfqewr}
\end{align}
Note that in the limit of $t\to\infty$ the error indeed converges to $\Gamma_-/\Gamma$ with an exponential rate given by $\kappa\Gamma$, where $\kappa>0$ can be a dimensionless constant to adjust the effective decay rate in a model where the effective decay rate $\kappa\Gamma$ does not match with the steady state error $\Gamma_-/\Gamma$ (see Eqn.\ (\ref{fixkappasdfqewrasdf}) and below). Since we are interested in the regime of small stationary error $\Gamma_-/\Gamma$, we can approximate and continue:
\begin{align}
E(t)~&\approx~\frac{\Gamma_-}{\Gamma}\,+\,e^{-t\kappa\Gamma}~=~\frac{\gamma^2}{g^2}\,\frac{h(N)}{f(N)}\,+\exp\left[-\gamma\,t\kappa\alpha^2f(N)\right]\label{erasdfwerasdfr}\\
&=~\left(\frac{\gamma}{g}\,\frac{\sqrt{h(N)}}{\sqrt{f(N)}}\right)^2\,+\,\exp\left[-\left(\frac{\gamma}{g}\,\frac{\sqrt{h(N)}}{\sqrt{f(N)}}\right)\,\left(t\kappa g\alpha^2\frac{f(N)^{3/2}}{h(N)^{1/2}}\right)\right]\\
&=~c^2\,+\,e^{-c\tau}~\,,\label{afterabbreviatingdecay}
\end{align}
where we abbreviate with a constant $c$ and a ``rescaled time'' $\tau$ as follows:
\begin{align}
c~&:=~\frac{\gamma}{g}\,\frac{\sqrt{h(N)}}{\sqrt{f(N)}}~,\label{candgammarelation}\\
\tau~&:=~t\kappa g\alpha^2\frac{f(N)^{3/2}}{h(N)^{1/2}}~.\label{physicaltandtaurelation}
\end{align}
We can treat $c$ as a free optimization variable, since $\gamma$ is a freely adjustable parameter, and thus we can adjust $c$ to any non-negative real number by choosing $\gamma$ appropriately (even if $g$ and $h(N)$ and $f(N)$ are fixed). Furthermore, $\tau$ is essentially the same as the physical time $t$, but rescaled by a fixed number (which in particular depends on $N$ and $g$).

\medskip

Now the dynamical problem is as follows: Given any fixed \emph{target error} ${\mathcal E}\in(0,1)$, we would like to find $c>0$ such that the time $\tau$ needed to achieve this error by Eqn.\ (\ref{afterabbreviatingdecay}) is minimized. Obviously, from (\ref{afterabbreviatingdecay}), any such suitable $c$ satisfies $c\in(0,\sqrt{{\mathcal E}})$. Thus, we can explicitly solve Eqn.\ (\ref{afterabbreviatingdecay}) for $\tau$ given $c$ and ${\mathcal E}$:
\begin{align}
\tau=\tau(c)~=~\frac{-\log({\mathcal E}-c^2)}{c}\qquad(c\in(0,\sqrt{{\mathcal E}}))~.\label{tauasafunctionofc}
\end{align}
To minimize this (rescaled) time $\tau=\tau(c)$, we set its derivative equal to zero (note that a minimum exists since $\lim_{c\to0}\tau(c)=\lim_{c\to\sqrt{{\mathcal E}}}\tau(c)=+\infty$):
\begin{align}
&\frac{d\tau(c)}{dc}~=~\frac{2}{{\mathcal E}-c^2}+\frac{\log({\mathcal E}-c^2)}{c^2}~=~0\label{derivativeoftauequalto0}\\
\Leftrightarrow\quad&-\log({\mathcal E}-c^2)~=~-2+\frac{2{\mathcal E}}{{\mathcal E}-c^2}\\
\Leftrightarrow\quad&\frac{1}{{\mathcal E}-c^2}~=~\exp\left[-2+\frac{2{\mathcal E}}{{\mathcal E}-c^2}\right]\\
\Leftrightarrow\quad&\frac{-2{\mathcal E}}{{\mathcal E}-c^2}\exp\left[-\frac{2{\mathcal E}}{{\mathcal E}-c^2}\right]~=~-2{\mathcal E}e^{-2}\\
\Leftrightarrow\quad&\frac{-2{\mathcal E}}{{\mathcal E}-c^2}~=~W(-2{\mathcal E}/e^2)~,\label{firstappearanceofLambertW}
\end{align}
where $W$ is (a branch of) the \emph{Lambert W function}, which satisfies $W(z)e^{W(z)}=z$ ($W$ is \emph{defined} to be a solution to this equation for every $z\in\mathbb C$). Now we have to identify the correct branch of $W(z)$ for our purposes: First, since ${\mathcal E}\in(0,1)$, the argument $-2{\mathcal E}/e^2$ in (\ref{firstappearanceofLambertW}) satisfies $-2{\mathcal E}/e^2\in(-2/e^2,0)\subseteq[-1/e,0)$; secondly, since $c\in(0,\sqrt{{\mathcal E}})$, the image in Eq.\ (\ref{firstappearanceofLambertW}) satisfies $-2{\mathcal E}/({\mathcal E}-c^2)\in(-\infty,-2)$. Both these things together mean that the correct branch (solution) of the function $W(z)$ in our problem is the branch $W_{-1}:[-1/e,0)\to(-\infty,-1],\,z\mapsto W_{-1}(z)$.

Then we can continue from (\ref{firstappearanceofLambertW}), and solve explicitly for the time-optimal $c$:
\begin{align}
c~=~\sqrt{{\mathcal E}}\,\sqrt{1+\frac{2}{W_{-1}(-2{\mathcal E}/e^2)}}~.\label{firstsolveforc}
\end{align}
The optimal time $\tau$ can be obtained by plugging this expression into (\ref{tauasafunctionofc}) and simplifying the expression, but there is a less direct and somewhat easier way: Observe from Eqs.\ (\ref{tauasafunctionofc}) and (\ref{derivativeoftauequalto0}) that
\begin{align}
\tau~&=~\frac{2c}{{\mathcal E}-c^2}~=~\frac{c}{{\mathcal E}}\,\frac{2{\mathcal E}}{{\mathcal E}-c^2}~\,\stackrel{[\text{Eq.}\ (\ref{firstappearanceofLambertW})]}{=}~\,\frac{-W_{-1}(-2{\mathcal E}/e^2)}{{\mathcal E}}\,c~\,\stackrel{[\text{Eq.}\ (\ref{firstsolveforc})]}{=}\\
&=\frac{1}{\sqrt{{\mathcal E}}}\sqrt{{W_{-1}(-2{\mathcal E}/e^2)^2+2W_{-1}(-2{\mathcal E}/e^2)}}\label{tauexactwithlambertasdfqwer}\\
&\approx\frac{1}{\sqrt{{\mathcal E}}}\log\frac{1}{{\mathcal E}}\qquad(\text{as}~\mathcal{E}\to0),
\end{align}
where in the last step we used the asymptotic approximation of our branch of the Lambert W function: $W_{-1}(x)=\log(-x)-\log(-\log(-x))+\mathcal{O}(1)$ as $x\to0$. This is justified when we are interested in the case of very small or asymptotically vanishing error ${\mathcal E}\to0$.

\medskip

Finally, we can use the above expressions to solve Eqs.\ (\ref{candgammarelation}) and (\ref{physicaltandtaurelation}) for the physically interesting optimal parameters $\gamma=\gamma(N,{\mathcal E})$ and  $t_{\rm GHZ}=t(N,{\mathcal E})$ using the values given in Eqs.\ (\ref{firstsolveforc}) and (\ref{tauexactwithlambertasdfqwer}). When we use the choices for $f(N)$ and $h(N)$ given in Eqs.\ (\ref{choiceforhNasdfqwer})  and (\ref{choiceforfNasdfqwer}), we obtain:
\begin{align}
\gamma~&=~gc\frac{f(N)^{1/2}}{h(N)^{1/2}}~=~\frac{4}{3\sqrt{3}}\,\frac{g\sqrt{{\mathcal E}}}{\sqrt{N}\log N}\,\,\sqrt{1+\frac{2}{W_{-1}(-2{\mathcal E}/e^2)}}\,\left(1+ \frac{5}{9\log^2N}\right)^{-1/2}\left(1+\frac{1}{3\log N}\right)^{-1/2}\label{exactformdynamicalproblemgammawithE}\\
&\approx~\frac{4}{3\sqrt{3}}\,\frac{g\sqrt{{\mathcal E}}}{\sqrt{N}\log N}~\qquad(\text{as}~N\to\infty,\,{\mathcal E}\to0),
\label{EqDynGamma}
\end{align}
Note that for small dynamical error $\mathcal{E}$ we have $W_{-1} \rightarrow 0$, such that Eq. \eqref{EqDynGamma} approaches the previous result in Eq. \eqref{gammawithXandZpumpingasdfqer}. This is due to the fact that for small $\mathcal{E}$ the stationary part of the error dominates.
The GHZ preparation time (defined as the time to reach a GHZ error of value $\mathcal{E}$) is:
\begin{align}
t_{\rm GHZ}~&=~\frac{\tau}{g\kappa\alpha^2}\frac{h(N)^{1/2}}{f(N)^{3/2}}~=~\frac{27\sqrt{3}}{8\kappa}\,\frac{\sqrt{N}\log^2N}{\alpha^2g\sqrt{{\mathcal E}}}\sqrt{W_{-1}^2\left(-\frac{2{\mathcal E}}{e^2}\right)+2W_{-1}\left(-\frac{2{\mathcal E}}{e^2}\right)}\,\left(1+ \frac{5}{9\log^2N}\right)^{1/2}\left(1+\frac{1}{3\log N}\right)^{3/2}\label{asdfqweryxcfvasdfqwer}\\
&\approx~\frac{27\sqrt{3}}{8\kappa}\,\frac{\sqrt{N}\log^2N}{\alpha^2g\sqrt{\mathcal E}}\,\log\frac{1}{{\mathcal E}}\qquad(\text{as}~N\to\infty,\,{\mathcal E}\to0)\,.\label{asdfqweryxcfv}
\end{align}

The approximations of the Lambert W function used above, i.e.\ $\sqrt{1+2/W_{-1}}\to1$ and $\sqrt{W^2_{-1}+2W_{-1}}\to\log(1/{\mathcal E})$, are good only for quite small ${\mathcal E}$. For ${\mathcal E}=0.1$, one should instead use the exact value $W_{-1}(-2{\mathcal E}/e^2)=-5.27$, leading to $\sqrt{1+2/W_{-1}}=0.788$ and $\sqrt{W^2_{-1}+2W_{-1}}=4.15$, which makes that Eq. \eqref{asdfqweryxcfv} is by a factor $1.8$ lower than Eq. \eqref{asdfqweryxcfvasdfqwer}. For ${\mathcal E}=0.03$ the corresponding value is $W_{-1}(-2{\mathcal E}/e^2)=-6.72$, leading to $\sqrt{1+2/W_{-1}}=0.838$ and $\sqrt{W^2_{-1}+2W_{-1}}=5.63$, which makes that Eq. \eqref{asdfqweryxcfv} is by a factor $1.6$ lower than Eq. \eqref{asdfqweryxcfvasdfqwer}. We can also find the relation between the stationary error $E=\Gamma_-/\Gamma$ (from Eq.\ (\ref{effectivedecayindynamicalproblemasdfqewr})) and the dynamical error ${\mathcal E}$:
\begin{align}
\frac{E}{{\mathcal E}}~&=~\frac{\Gamma_-}{{\mathcal E}\,\Gamma}~\stackrel{[\text{Eqs.}\ (\ref{Gammaindynamicalsectionasdf})-(\ref{choiceforfNasdfqwer})]}{=}~\frac{\gamma^2}{{\mathcal E}\,g^2}\,\frac{3N\log N}{8}\left(1+\frac{5}{9\log^2N}\right)\,\frac{9\log N}{2}\,\left(1+\frac{1}{3\log N}\right)\\
&\stackrel{[\text{Eq.}\ (\ref{exactformdynamicalproblemgammawithE})]}{=}~~1+\frac{2}{W_{-1}(-2{\mathcal E}/e^2)}\,,\label{relationstatanddynerror}
\end{align}
which is \emph{in}dependent of $N$. Thus, using the above values, for ${\mathcal E}=0.1$ we get for the stationary error at the optimal parameters $E=0.62{\mathcal E}$, whereas for ${\mathcal E}=0.03$ we get $E=0.70{\mathcal E}$.

It now remains to fix the value of $\kappa$ (appearing in (\ref{asdfqweryxcfvasdfqwer})--(\ref{asdfqweryxcfv})) in an appropriate way, namely such that the model (\ref{effectivedecayindynamicalproblemasdfqewr}) matches the GHZ preparation process from Sections \ref{subsectioncompartmentmodel} and \ref{subsectiontransitionmatrixetc} as well as possible. For this, note that the effective GHZ preparation rate $\Gamma_{\rm GHZ}$ from Section \ref{subsectiontransitionmatrixetc} can be inferred from Eq.\ (\ref{firsteqnasdfqweryxcv}):
\begin{align}
\kappa\Gamma~\stackrel{!}{=}~\Gamma_{\rm GHZ}~\equiv~\frac{1}{\tau_{\rm GHZ}}~=~\frac{\Omega^2}{\gamma}\,\frac{2}{3\log N}\,\frac{-\log\left(1-P_4(t_0)\right)}{\Gamma_Z^+ t_0}\,.\label{fixkappasdfqewrasdf}
\end{align}
Using the value of $\Gamma$ from Eqs.\ (\ref{Gammaindynamicalsectionasdf}) and (\ref{choiceforfNasdfqwer}), we can solve for $\kappa$:
\begin{align}
\kappa~=~3\,\left(\frac{-\log(1-P_4(t_0))}{\Gamma^+_Z t_0}\right)\,\left(1+\frac{1}{3\log N}\right)\,.
\end{align}
Evaluating this as in Section \ref{subsectiontransitionmatrixetc}, we get the following table:
\begin{center}
\begin{tabular}{|l||l|l|l|l|l|l|l|l|l|l|l|}
\hline
$N$ & $2$ & $3$ & $4$ & $5$  & $6$ & $7$ & $8$ & $10$ & $20$ & $50$ & $100$  \\
\hline
$\kappa$ & 0.28 & 0.32 & 0.34 & 0.35 & 0.36 & 0.36 & 0.37 & 0.38 & 0.39 & 0.40 & 0.41\\
\hline
\end{tabular}
\end{center}

\subsection{GHZ scaling analysis for strong driving}
\label{SecGHZStrong}

When taking power broadening into account (see Section \ref{powerbroadeningdescriptionsubsection}), then instead of Eq.\ (\ref{rateasdfasdfasdfabasdbasdbasb}) from the weak driving scenario, the favorable transition rates of the Z process are now given by Eq.\ (\ref{GammaGoodPower}) (for $n_1=1,2,\ldots,N-1$):
\begin{align}
\Gamma_{n_1 \rightarrow n_1 - 1, \gamma 0, Z}
~=~\frac{2 n_1 \gamma_{0e} (\Omega_Z^{(F=n_1)})^2}{(\gamma_e + \kappa_b)^2 + 2 n_1 (\Omega_Z^{(F=n_1)})^2}~=~\frac{\gamma F \Omega_{F}^2}{\gamma^2+2 F \Omega_{F}^2}\,,
\end{align}
where we have again used the parameter values and abbreviations $\gamma_e\equiv\gamma$, $\gamma_{0e}=\gamma/2$, $\Omega_Z^{(F=n_1)}=\Omega_{F}$ as in Section \ref{SecPreparation}.
Thus, instead of (\ref{worstcaseZpreptimewithfactor2}), the Z pumping time is now:
\begin{align}
\tau_{n_1=N-1 \rightarrow {\rm GHZ}} &= 2 \tau_{n_1=N-1 \rightarrow n_1=0} = 2\sum_{n_1=1}^{N-1}(\Gamma_{n_1 \rightarrow n_1 - 1, \gamma 0, Z})^{-1}\\
&=2\sum_{F=1}^{N-1}\left(\frac{\gamma}{\Omega_{F}^2 F}+\frac{2}{\gamma}\right)=\frac{4(N-1)}{\gamma}\,+\,\sum_{F=1}^{N-1}\frac{2\gamma}{F \Omega_F^2}\label{tau1NGHZpowerasdf}
\end{align}
The error rate from (\ref{restateGammaminus}) remains unchanged:
\begin{align}
\Gamma_{{\rm GHZ} \rightarrow ?, \gamma, Z} ~=~\frac{3 \gamma}{16 g^2} N \sum_{F=1}^{N-1}\frac{F \Omega_{F}^2}{(N - F)^2}.\label{GammaminusGHZpowerasdfqwer}
\end{align}
As below (\ref{asdfasdfzuio}), we now minimize $\Gamma_{{\rm GHZ} \rightarrow ?, \gamma, Z}$ (as a function of the variables $\{\Omega_f\}$) while keeping $\tau_{n_1=N-1 \rightarrow {\rm GHZ}}$ constant. This leads, by the method of Lagrange multipliers, to:
\begin{align}\label{OmegafGHZpowerbroaderninglambda}
\Omega_F^2~=~\lambda\frac{N-F}{F}\qquad(\text{for}~F=1,2,\ldots,N-1)\,,
\end{align}
where $\lambda$ is a Lagrange multiplier (that has units of frequency$^2$), which we will determine later. Plugging this back into (\ref{tau1NGHZpowerasdf}) and (\ref{GammaminusGHZpowerasdfqwer}), we get:
\begin{align}\label{tau1NGHZpowerwithlambda}
\tau_{N-1\to {\rm GHZ}}~=~\frac{4(N-1)}{\gamma}\,+\,\frac{2\gamma}{\lambda}\sum_{F=1}^{N-1}\frac{1}{N-F}~\simeq~\frac{4(N-1)}{\gamma}\,+\,\frac{2\gamma}{\lambda}\log N\,,
\end{align}
\begin{align}
\Gamma_{{\rm GHZ} \to ?,\gamma, Z}~=~\frac{3\gamma\lambda}{16g^2}N\log N\,.
\end{align}
Thus, the stationary Z error is (cf.\ (\ref{errorinZpumping})):
\begin{align}
E_Z~=~(\Gamma_{{\rm GHZ} \to ?, \gamma, Z})(\tau_{N-1\to {\rm GHZ}})~=\frac{3\lambda}{4g^2}(N-1)N\log N\,+\,\frac{3\gamma^2}{8g^2}N\log^2N\,.
\end{align}
Thus, when the desired stationary Z error $E_Z$ is given, then $\lambda$ and $\gamma$ are determined by each other (since $g$ and $N$ are fixed):
\begin{align}
\lambda~=~\frac{8g^2E_Z-3\gamma^2N\log^2N}{6(N-1)N\log N}\qquad\left(\text{for}~~\gamma^2\in\left[0,\frac{8g^2E_Z}{3N\log^2N}\right]\right)\,.\label{lambdaGHZlangranemultiplier}
\end{align}
Plugging this back into (\ref{tau1NGHZpowerwithlambda}), we get:
\begin{align}
\tau_{N-1\to {\rm GHZ}}~=~4(N-1)\left[\frac{1}{\gamma}\,+\,\frac{\gamma}{\frac{8g^2E_Z}{3N\log^2N}-\gamma^2}\right]\,.
\end{align}
For fixed $N$, $g$, $E_Z$, and $\gamma$, this is the minimal Z pumping time (i.e.\ minimized over all choices of pumping rates $\{\Omega_f\}$). Since we do not want to fix $\gamma$ a priori, we minimize the last expression over $\gamma$, finding that the optimal choice is
\begin{align}\label{gammainGHZsectionwithoutpower broadening}
\gamma~=~\sqrt{\frac{8g^2E_Z}{9N\log^2N}}\,.
\end{align}
The corresponding pumping strengths can now be computed from (\ref{OmegafGHZpowerbroaderninglambda}) and (\ref{lambdaGHZlangranemultiplier}):
\begin{align}\label{Omegafrateswithpower broadening}
\Omega_F~=~\sqrt{\frac{8g^2E_Z}{9N(N-1)\log N}\,\frac{N-F}{F}}\qquad(\text{for}~F=1,\ldots,N-1)\,,
\end{align}
and the optimal Z pumping time for given $N$, $g$, $E_Z$ is then (compare to (\ref{tau1Nforoptimalconsteta2}) without power broadening):
\begin{align}
\tau_{N-1\to {\rm GHZ}}~=~\frac{9}{\sqrt{2}}\,\frac{(N-1)\sqrt{N}\log N}{g\sqrt{E_Z}}~\simeq~\frac{9}{\sqrt{2}}\,\frac{N^{3/2}\log N}{g\sqrt{E_Z}}\,.
\end{align}
From now on we set $(N-1)\simeq N$, as this neglects only subleading terms.

\bigskip

We also take power broadening into account for the desired X pumping rate (see (\ref{goodXratepowerbroadening})) and for the X toss rate (see \ref{tossXratepowerboradeningasdfqwerasdf}), which we both compute as in (\ref{startcomputinggoodXrateasdfqwerasdf})--(\ref{goodXrate2}) and (\ref{Xtossrate2}), and we take the detrimental X rate from (\ref{badXrate}):
\begin{align}
\Gamma_X^+&~=~\sum_{n_-=1,3,...}^N \frac{1}{2^{N-1}}   {N \choose n_-}\frac{2 \gamma_f}{(\gamma_f + \kappa_c)^2 + 2 n_- (\Omega_X^{(F=n_-)})^2} {n_- (\Omega_X^{(F=n_-)})^2}\\
&~\simeq~\frac{2 \gamma_f}{\gamma_f^2 + 2 (N/2)\Omega^2} {\frac{N}{2} \Omega^2}~=~N\frac{\gamma_f \Omega^2}{\gamma_f^2+N\Omega^2}\,,\\
\Gamma_X^{toss}&~=~\frac{1}{2}\Gamma_X^+~\simeq~\frac{N}{2}\frac{\gamma_f \Omega^2}{\gamma_f^2+N\Omega_X^2}\,,\\
\Gamma_X^-&~=~\frac{\gamma_f \Omega^2}{g^2}\,\frac{5N^2}{16}\,.
\end{align}

Now we adjust the X parameters $\Omega$ and $\gamma_f$ such that the X rates agree with the corresponding Z rates (cf.\ Section \ref{subsectioncompartmentmodel}), i.e.\ $\Gamma_X^+\equiv2\Gamma_X^{toss}=\Gamma_Z^+:=1/(\tau_{N-1\to {\rm GHZ}})=\frac{\sqrt{2}}{9}\frac{g\sqrt{E_Z}}{N^{3/2}\log N}$ and $\Gamma_X^-=\Gamma_{{\rm GHZ} \to ?}^-=\frac{\sqrt{2}}{9}\frac{g E_Z^{3/2}}{N^{3/2}\log N}$. To solve this for $\gamma_f$ and $\Omega$ exactly, one would have to solve a cubic equation. As this is quite cumbersome and uninformative, we are looking for solutions which satisfy the following scaling ansaetze: $\gamma_f\simeq N^\alpha(\log N)^\beta$ and $\Omega\simeq N^\phi(\log N)^\psi$. Then, one finds actually \emph{two} possible solutions for $\Omega$ and $\gamma_f$, which lead to the desired scaling of $\Gamma_X^+$ and $\Gamma_X^-$ (as $N$ becomes large). One of the solutions is given by
\begin{align}\label{twodifferentchoicesforgammaOmegaasdfqwer}
\Omega~&=~\frac{2^{5/4}}{3\cdot5^{1/4}}\frac{gE_Z^{1/2}}{N^{3/2}(\log N)^{1/2}}
\\
\gamma_f~&=~\frac{4}{\sqrt{5}}\frac{gE_Z^{1/2}}{N^{1/2}}.
\end{align}
Formally, another solution exists, but this is in the extremely saturated regime where our effective operators do not apply. 

\bigskip

Finally, we need to find the relation between the above rate $\Gamma_Z^+:=1/(\tau_{N-1\to {\rm GHZ}, Z})$ and the total GHZ preparation time $\tau_{\rm GHZ} \equiv 1/\Gamma_+$ (which includes the errors and X toss) and the total stationary error $E$, just as in Section \ref{subsectiontransitionmatrixetc}. For this, we consider a simplified 3-compartment model which constitutes a very good approximation in the large-$N$-limit, in which the parameters (\ref{twodifferentchoicesforgammaOmegaasdfqwer}) were computed in the first place. Here, compartment A comprises the states with $n_1=1,2,\ldots,N-1$, compartment B is the state $\ket{\rm GHZ_-}$ and compartment C the state $\ket{\rm GHZ}$. Then the transition matrix is (cf.\ (\ref{transitionmatrix4x4pess}) and Fig.\ \ref{comp4pessfig}):
\begin{align}
T~=~\begin{pmatrix}-2\Gamma_Z^+-\Gamma_X^{toss}&\Gamma^+_X&\Gamma_-&\\\Gamma_Z^++\Gamma_X^{toss}&-\Gamma_X^+&0\\\Gamma_Z^+&0&-\Gamma_-\end{pmatrix}~
=~\Gamma_Z^+\begin{pmatrix}-5/2&1&\Gamma_-/\Gamma_Z^+&\\3/2&-1&0\\1&0&-\Gamma_-/\Gamma_Z^+\end{pmatrix}~,
\end{align}
again with $\Gamma_-=\Gamma_X^-+\Gamma_Z^-$, and again as in Section \ref{subsectiontransitionmatrixetc} we denote by $T_+$ the transition matrix without the ``bad'' rates $\Gamma_-$. From the stationary vector $p_\infty$ satisfying $Tp_\infty=0$ we can again compute the total stationary error $E$:
\begin{align}
E=1-\frac{1}{1+\frac{5}{2}\frac{\Gamma_-}{\Gamma_Z^+}}\simeq\frac{5}{2}\frac{\Gamma_-}{\Gamma_Z^+}=5E_Z.
\end{align}
We therefore have to set $E_Z=E/5$ in all previous expressions. The relation between $\Gamma_Z^+$ and $\Gamma_+\equiv1/\tau_{\rm GHZ}$ is similar to Eq.\ (\ref{rateasdfqwerycxvasdfqutqw}):
\begin{align}
\Gamma_+~=~\Gamma_Z^+\,\frac{-\log\left(1-P_C(t_0)\right)}{\Gamma_Z^+t_0}\,,
\end{align}
where again $P_C$ denotes the population in the GHZ state when starting from $\ket{\rm GHZ_-}$ state. Computing this for the above transition matrix and plugging in the values for $\Gamma_Z^+$ and $E$ from above, we obtain:
\begin{align}
\Gamma_+~=~0.216\cdot\Gamma_Z^+~=~0.0339\cdot\frac{g\sqrt{E_Z}}{N^{3/2}\log N}~=~0.0152\cdot\frac{g\sqrt{E}}{N^{3/2}\log N}\,.
\end{align}
Thus, in the large-$N$ limit, the final GHZ preparation time $\tau_{\rm GHZ}$ is:
\begin{align}
\tau_{\rm GHZ}~&=~\frac{1}{\Gamma_+}~\approx~66\frac{N^{3/2}\log N}{g\sqrt{E}}\,.
\label{finalGHZpowerasdfqwerycxv}
\end{align}

This is achieved by the following parameter choices in terms of $g$, $N$, $E$:
\begin{align}
\gamma~&=~0.42\cdot{\frac{g\sqrt{E}}{\sqrt{N}\log N}}\qquad\text{(this is the $\gamma$-rate for Z pumping)}\,,\\
\Omega_F~&=~0.42\cdot{\frac{g\sqrt{E}}{N\sqrt{\log N}}}\sqrt{\frac{N-F}{F}}\qquad(\text{for}~F=1,\ldots,N-1)\,,
\\
\Omega~&=~0.24\cdot\frac{gE^{1/2}}{N^{3/2}(\log N)^{1/2}}
\\
\gamma_f~&=~0.80\cdot\frac{gE^{1/2}}{N^{1/2}}
.
\end{align}
Again, as in Eq.\ (\ref{relationstatanddynerror}), if one is interested in the dynamical error ${\mathcal E}$ instead of the static error $E$, one should everywhere set $E=0.62{\mathcal E}$ for ${\mathcal E}=0.1$ (or $E=0.70{\mathcal E}$ for ${\mathcal E}=0.03$). Furthermore, the GHZ preparation time is then prolonged by an additional factor of $\log(1/{\mathcal E})$ (see Eq.\ (\ref{asdfqweryxcfv})).

\subsection{Dynamical problem for fixed preparation time $T$}
If the maximal preparation time $T=t_{max}$ of the preparation procedure is limited, e.g.\ by experimental constraints or by the available coherence times of the underlying hardware, we can still try to adjust the remaining parameters of the preparation scheme in such a way, that the error ${\mathcal E}=E(T)=1-F(T)$ after this fixed time $T$ is minimized. We first take the same setup of the dynamical problem as in Subsection \ref{AppDynamical}, i.e.\ as in Eqs.\ (\ref{Gammaindynamicalsectionasdf})-(\ref{Gammaminuswithh}) and (\ref{erasdfwerasdfr}):
\begin{align}
\Gamma~&=~\frac{\Omega^2}{\gamma}f(N)~=~\alpha^2\gamma f(N)~\label{asdfasdfasdfasdfasdfasdfasdfasdf},\\
\Gamma_-~&=~\frac{\gamma\Omega^2}{g^2}h(N)~,\\
E(T)~&\approx~\frac{\Gamma_-}{\Gamma}\,+\,e^{-T\kappa\Gamma}~=~\frac{\gamma^2}{g^2}\,\frac{h(N)}{f(N)}\,+\exp\left[-\gamma\,T\kappa\alpha^2f(N)\right]\,,\label{asdfqewrasdfjknkjhhjhj}
\end{align}
where we again assumed a relation $\Omega=\alpha\gamma$ between $\Omega$ and $\gamma$ with a fixed parameter $\alpha$. It is now our task to minimize the preparation error $E(T)$ for each fixed $T$, $N$, $g$, and for the given constants and functions $\alpha$, $\kappa$, $f(N)$, and $h(N)$. The free parameter is thus $\gamma$ (which determines $\Omega=\alpha\gamma$).

This optimization of the error $E(T)$ for fixed preparation time $T$ can be done similarly as the optimization in Subsection \ref{AppDynamical}. To find the optimal parameters $\gamma$ and $\Omega$, we demand $\frac{d}{d\gamma}E(T)=0$, which becomes:
\begin{align}
2\frac{\gamma}{g^2}\frac{h(N)}{f(N)}-T\kappa\alpha^2f(N)e^{-\gamma T\kappa\alpha^2f(N)}=0\,.
\end{align}
This is equivalent to
\begin{align}
(\gamma T\kappa\alpha^2 f(N))\,e^{\gamma T\kappa\alpha^2f(N)}=\frac{T^2\kappa^2\alpha^4f^3(N)g^2}{2h(N)}\,,\label{bringtolambertWform}
\end{align}
whose solution is again given by the Lambert W function:
\begin{align}
\gamma T\kappa\alpha^2 f(N)=W_0\left(\frac{T^2\kappa^2\alpha^4f^3(N)g^2}{2h(N)}\right)\,,\label{blasdfqewradsfb}
\end{align}
where this time we have to take the branch $W_0:[0,\infty)\to[0,\infty), z\mapsto W_0(z)$ since both sides are nonnegative. We can now solve the condition (\ref{bringtolambertWform}) for the optimal $e^{-\gamma T\kappa\alpha^2f(N)}$ and plug this back into the optimal preparation error $E(T)$ from (\ref{asdfqewrasdfjknkjhhjhj}) together with the solution in terms of the Lambert W function.

Then we obtain for the optimal preparation error $E(T)$ after a fixed time $T$:
\begin{align}
E(T)=\frac{h(N)}{f^3(N)}\,\frac{W(W+2)}{T^2\kappa^2\alpha^4g^2}\,,\qquad\text{where}~~W=W_0\left(\frac{T^2\kappa^2\alpha^4f^3(N)g^2}{2h(N)}\right)\,.
\end{align}
and the optimal rate $\gamma$ is determined by (\ref{blasdfqewradsfb}). As we have seen previously, plausible behaviours of the underlying rates $\Gamma$ and $\Gamma_-$ are given by a $N$-independent $f(N)\sim1$ and a linear $h(N)\sim N$. The expression for the optimal error $E(T)$ at fixed time $T$ and particle number $N$ can then be evaluated numerically. Unlike in Subsection \ref{AppDynamical} however, an asymptotic formula for the Lambert W function cannot be applied here since the optimal solution lies right in the middle of the competing polynomially growing (in $\gamma$) and exponentially decaying behaviours in (\ref{asdfqewrasdfjknkjhhjhj}).

\bigskip

To obtain more analytical insight, we will thus now upper-bound the particle number $N$, for a given fixed preparation time and prescribed maximal error ${\mathcal E}$. For concreteness of the exposition we will also take $f(N)=1$ and $h(N)=N$ in (\ref{asdfasdfasdfasdfasdfasdfasdfasdf})-(\ref{asdfqewrasdfjknkjhhjhj}), although the computation is easily generalized to more general behaviours of $\Gamma$ and $\Gamma_-$. If we demand $E(T)\leq{\mathcal E}$, then from (\ref{asdfqewrasdfjknkjhhjhj}) we certainly need to have
\begin{align}
&\frac{\Gamma_-}{\Gamma}=N\frac{\gamma^2}{g^2}~\leq~{\mathcal E}\\
\text{and}\quad&e^{-T\kappa\Gamma}=e^{-\gamma T\kappa\alpha^2}~\leq~{\mathcal E}\,.
\end{align}
From the last inequality, we obtain the condition $\gamma T\kappa\alpha^2\geq\log(1/{\mathcal E})$, i.e.\ the requirement $1/\gamma\leq T\kappa\alpha^2/\log(1/{\mathcal E})$ on $\gamma$. Using this in the first of the two inequalities, we finally obtain:
\begin{align}
N~\leq~{\mathcal E}\,\frac{g^2}{\gamma^2}~\leq~T^2\kappa^2\alpha^4g^2\,\frac{{\mathcal E}}{\log^2(1/{\mathcal E})}~=~\kappa^2\,(gT)^2\,\left(\frac{\Omega}{\gamma}\right)^4\,\frac{{\mathcal E}}{\log^2(1/{\mathcal E})}~\sim~(gT)^2\,\frac{{\mathcal E}}{\log^2(1/{\mathcal E})}\,.\label{fiadfqewrxcvasdfqwer}
\end{align}
Thus, if the maximal preparation time $T$ is given and a certain maximal error ${\mathcal E}$ is to be achieved, then the maximal size of the GHZ state is limited to a number $N$ of particles that is proportional to the square of the ``dimensionless coherence time'' $gT$ and also proportional to the allowed error ${\mathcal E}$ (up to logarithmic factors).

\bigskip

One can notice that the bound on $N$ from (\ref{fiadfqewrxcvasdfqwer}) could also be simply obtained (up to logarithmic factors) by solving the main result (\ref{asdfqweryxcfv}) from Subsection \ref{AppDynamical} for $N$. Thus, if we now want to take the effect of powerbroadening into account, we solve the main final result from Section \ref{SecGHZStrong} for $N$, namely Eq.\ (\ref{finalGHZpowerasdfqwerycxv}), to obtain (up to logarithmic factors):
\begin{align}
N~\lesssim~\frac{1}{16}\,(gT)^{2/3}\,E^{1/3}\,.
\end{align}
Note that this same result would be obtained from the above discussion by a choice of $f(N)$ and $h(N)$ satisfying $h^{1/3}(N)/f(N)\sim N$, i.e.\ for example by rates $\Gamma$ and $\Gamma_-$ given as $\Gamma=\Omega^2/\gamma$ and $\Gamma_-=N^3\gamma\Omega^2/g^2$.


\begin{thebibliography}{99}

\bibitem{NielsenChuang}
M. A. Nielsen and I. L. Chuang,
\textit{Quantum computation and quantum information} (Cambridge University Press, Cambridge, 2000).

\bibitem{ShorErrorCorrection}
P. W. Shor,
\textit{Phys. Rev. A} \textbf{52}, 2493 (1995).

\bibitem{Steane}
A. M. Steane,
\textit{Phys. Rev. Lett.} \textbf{77}, 793 (1996).

\bibitem{Fleischhauer}
M. Fleischhauer and M. D. Lukin,
Phys. Rev. A \textbf{65}, 022314 (2002).

\bibitem{Giovanetti}
V. Giovannetti, S. Lloyd, and L. Maccone,
Science \textbf{306}, 1330 (2004).

\bibitem{Toth}
G. Toth and I. Apellaniz,
J. Phys. A: Math. Theor. \textbf{47}, 424006 (2014).

\bibitem{Rauschenbeutel}
A. Rauschenbeutel \textit{et al.},
Science \textbf{288}, 2024 (2000).

\bibitem{Sackett}
C. A. Sackett \textit{et al.},
Nature \textbf{404}, 256 (2000).

\bibitem{SchmidtKaler}
F. Schmidt-Kaler, H. H\"{a}ffner, M. Riebe, S. Gulde, G. P. T. Lancaster, T. Deuschle, C. Becher, C. F. Roos, J. Eschner, and R. Blatt,
Nature \textbf{422}, 408 (2003).

\bibitem{DiCarloNeeley}
L. DiCarlo \textit{et al.},
Nature \textbf{467}, 574 (2010);
M. Neeley \textit{et al.},
Nature \textbf{467}, 570 (2010).

\bibitem{Leibfried}
D. Leibfried \textit{et al.},
Nature \textbf{438}, 639 (2005).

\bibitem{Haffner}
H. H\"{a}ffner \textit{et al.},
Nature \textbf{438}, 643 (2005).

\bibitem{Monz}
T. Monz, P. Schindler, J. T. Barreiro, M. Chwalla, D. Nigg, W. A. Coish, M. Harlander, W. H\"{a}nsel, M. Hennrich, and R. Blatt,
Phys. Rev. Lett. \textbf{106}, 130506 (2011).

\bibitem{WisemanMilburn}
H. M. Wiseman and G. J. Milburn, 
\textit{Quantum Measurement and Control} (Cambridge University Press, Cambridge 2009).

\bibitem{Bergholm}
V. Bergholm and T. Schulte-Herbrueggen,
arXiv:1206.4945 (2012).

\bibitem{Riste}
D. Riste, M. Dukalski, C. A. Watson, G. de Lange, M. J. Tiggelman, Y. M. Blanter, K. W. Lehnert, R. N. Schouten, and L. DiCarlo,
Nature \textbf{502}, 350 (2013).

\bibitem{DiVincenzo}
D. P. DiVincenzo, 
Fortschr. Phys. \textbf{48}, 771 (2000).

\bibitem{Ladd}
T. D. Ladd, F. Jelezko, R. Laflamme, Y. Nakamura, C. Monroe, and J. L. O'Brien,
Nature \textbf{464}, 45 (2010).

\bibitem{Beige1}
A. Beige,
Phys. Rev. A \textbf{69}, 012303 (2004).

\bibitem{Kraus}
B. Kraus, H. P. B\"{u}chler, S. Diehl, A. Kantian, A. Micheli, and P. Zoller,
Phys. Rev. A \textbf{78}, 042307 (2008).

\bibitem{Ticozzi1}
F. Ticozzi and L. Viola,
IEEE T. Automat. Contr. \textbf{53}, 2048 (2008).

\bibitem{Diehl}
S. Diehl, A. Micheli, A. Kantian, B. Kraus, H. P. B\"{u}chler, and P. Zoller,
Nat. Phys. \textbf{4}, 878 (2008).

\bibitem{VWC}
F. Verstraete, M. M. Wolf, and J. I. Cirac,
Nat. Phys. \textbf{5}, 633 (2009).

\bibitem{PCZ}
J. F. Poyatos, J. I. Cirac, and P. Zoller,
Phys. Rev. Lett. \textbf{77}, 4728 (1996).

\bibitem{PHBK}
M. B. Plenio, S. F. Huelga, A. Beige, and P. L. Knight,
Phys. Rev. A \textbf{59}, 2468 (1999). 

\bibitem{Beige2}
A. Beige, D. Braun and P. L. Knight,
New J. Phys. \textbf{2}, 22 (2000). 

\bibitem{WS}
X. T. Wang and S. G. Schirmer,
arXiv:1005.2114.

\bibitem{KRS}
M. J. Kastoryano, F. Reiter, and A. S. S\o{}rensen,
Phys. Rev. Lett. \textbf{106}, 090502 (2011).

\bibitem{Gullans}
M. Gullans, T. G. Tiecke, D. E. Chang, J. Chang, J. Feist, J. D. Thompson, J. I. Cirac, P. Zoller, and M. D. Lukin,
Phys. Rev. Lett. \textbf{109}, 235309 (2012).

\bibitem{Carr}
A. W. Carr and M. Saffman,
Phys. Rev. Lett. \textbf{111}, 033607 (2013).

\bibitem{Schuetz}
M. J. A. Schuetz, E. M. Kessler, L. M. K. Vandersypen, J. I. Cirac, and G. Giedke,
Phys. Rev. Lett. \textbf{111}, 246802 (2013).

\bibitem{Krauter}
H. Krauter, C. A. Muschik, K. Jensen, W. Wasilewski, J. M. Pedersen, J. I. Cirac, and E. S. Polzik,
Phys. Rev. Lett. \textbf{107}, 080503 (2011).

\bibitem{Barreiro}
J. T. Barreiro, M M\"{u}ller, P. Schindler, D. Nigg, T. Monz, M. Chwalla, M. Hennrich, C. F. Roos, P. Zoller, and R. Blatt,
Nature \textbf{470}, 486 (2011).

\bibitem{Lin}
Y. Lin, J. P. Gaebler, F. Reiter, T. R. Tan, R. Bowler, A. S. S\o{}rensen, D. Leibfried, and D. J. Wineland,
\textit{Nature} \textbf{504}, 415 (2013).

\bibitem{Shankar}
S. Shankar, M. Hatridge, Z. Leghtas, K. M. Sliwa, A. Narla, U. Vool, S. M. Girvin, L. Frunzio, M. Mirrahimi, and M. H. Devoret,
Nature \textbf{504}, 419 (2013).

\bibitem{SchneiderMilburn}
S. Schneider and G. J. Milburn,
Phys. Rev. A \textbf{65}, 042107 (2002).

\bibitem{Vacanti}
G. Vacanti and A. Beige,
New J. Phys. \textbf{11}, 083008 (2009).

\bibitem{Weimer}
H. Weimer, M. M\"{u}ller, I. Lesanovsky, P. Zoller, and H. P. B\"{u}chler,
Nat. Phys. \textbf{6}, 382 (2010).

\bibitem{Vollbrecht}
K. G. H. Vollbrecht, C. A. Muschik, and J. I. Cirac,
Phys. Rev. Lett. \textbf{107}, 120502 (2011).

\bibitem{CBK}
J. Cho, S. Bose, and M. S. Kim,
Phys. Rev. Lett. \textbf{106}, 020504 (2011).

\bibitem{Stevenson}
R. N. Stevenson, J. J. Hope, and A. R. R. Carvalho,
Phys. Rev. A \textbf{84}, 022332 (2011).

\bibitem{KRW}
M. J. Kastoryano, D. Reeb, and M. M. Wolf,
J. Phys. A \textbf{45}, 075307 (2012).

\bibitem{FossFeig}
M. Foss-Feig, A. J. Daley, J. K. Thompson, and A. M. Rey,
Phys. Rev. Lett. \textbf{109}, 230501 (2012).

\bibitem{Plasmons}
A. Gonzalez-Tudela and D. Porras,
Phys. Rev. Lett. \textbf{110}, 080502 (2013).

\bibitem{Cormick}
C. Cormick, A. Bermudez, S. F. Huelga, and M. B. Plenio,
New. J. Phys. \textbf{15}, 073027 (2013).

\bibitem{Honing}
M. H\"{o}ning, D. Muth, D. Petrosyan, and M. Fleischhauer,
Phys. Rev. A \textbf{87}, 023401 (2013).

\bibitem{LCC}
S. K. Lee, J. Cho, and K. S. Choi,
New J. Phys. \textbf{17}, 113053 (2015).

\bibitem{Ticozzi2}
F. Ticozzi and L. Viola,
Quantum Inf. Comput. \textbf{14}, 0265 (2014).

\bibitem{Rao}
D. D. B. Rao and K. M\o{}lmer,
Phys. Rev. A \textbf{90}, 062319 (2014).

\bibitem{Pichler}
H. Pichler, T. Ramos, A. J. Daley, and P. Zoller,
Phys. Rev. A \textbf{91}, 042116 (2015).

\bibitem{Lee}
T. E. Lee, F. Reiter, and N. Moiseyev,
Phys. Rev. Lett. \textbf{113}, 250401 (2014).

\bibitem{Happer}
W. Happer,
Rev. Mod. Phys. \textbf{44}, 169 (1972).

\bibitem{SI}
See Supplementary Information for the detailed analysis.

\bibitem{EO}
F. Reiter and A. S. S\o{}rensen,
Phys. Rev. A \textbf{85}, 032111 (2012).

\bibitem{Harty}
T. P. Harty, D. T. C. Allcock, C. J. Ballance, L. Guidoni, H. A. Janacek, N. M. Linke, D. N. Stacey, and D. M. Lucas,
Phys. Rev. Lett. \textbf{113}, 220501 (2014).

\bibitem{Banachiewicz} S. Puntanen and G. Styan,
\textit{The Schur Complement and Its Applications},
(Springer, New York, 2005).

\bibitem{boydconvex} S. Boyd and L. Vandenberghe,
\textit{Convex Optimization}
(Cambridge University Press, Cambridge, 2004)

\end{thebibliography}
\end{document}